\documentclass[12pt,preprintnumbers,amsmath,nofootinbib,secnumarabic]{revtex4-1}

\usepackage{graphicx}  
\usepackage{dcolumn}   
\usepackage{bm}        
\usepackage{graphicx}
\usepackage{caption}
\usepackage{subcaption}
\usepackage{epsfig,amsfonts,amsthm}
\usepackage{epstopdf}
\usepackage{mathrsfs}
\usepackage{dsfont}
\usepackage{pstricks}
\usepackage{color}
\usepackage{amssymb,amsmath,amsbsy,bbold}
\usepackage{latexsym}

\def\beq{\begin{equation}}
\def\eeq{\end{equation}}
\newenvironment{Eqnarray}%
     {\arraycolsep 0.14em\begin{eqnarray}}{\end{eqnarray}}
\def\beqa{\begin{Eqnarray}}
\def\eeqa{\end{Eqnarray}}
\def\lsim{\mathrel{\raise.3ex\hbox{$<$\kern-.75em\lower1ex\hbox{$\sim$}}}}
\def\gsim{\mathrel{\raise.3ex\hbox{$>$\kern-.75em\lower1ex\hbox{$\sim$}}}}

\def\anti{\overline}
\def\wtil{\widetilde}
\def\ur{U_R}
\def\dr{D_R}

\def\abar{{\bar a}}
\def\bbar{{\bar b}}
\def\cbar{{\bar c}}
\def\dbar{{\bar d}}

\def\qlo{Q^0_L}
\def\uro{U^0_R}
\def\dro{D^0_R}

\def\ur{U_R}
\def\dr{D_R}
\def\eiuo{\eta_1^{U,0}}
\def\eiiuo{\eta_2^{U,0}}
\def\eido{\eta_1^{D,0}}
\def\eiido{\eta_2^{D,0}}

\def\eiuoa{\eta_a^{U,0}}
\def\eidoa{\eta_a^{D,0}}

\def\eidoab{\eta_{\abar}^{D,0}}

\def\eiuob{\eta_b^{U,0}}
\def\eidob{\eta_b^{D,0}}
\def\eiuobb{\eta_{\bbar}^{U,0}}
\def\eidobb{\eta_{\bbar}^{D,0}}

\def\elo{E^0_L}

\def\ero{E^0_R}

\def\er{E_R}
\def\eieo{\eta_1^{E,0}}
\def\eiieo{\eta_2^{E,0}}

\def\eieoa{\eta_a^{E,0}}
\def\eieoab{\eta_{\abar}^{E,0}}

\def\eieob{\eta_b^{E,0}}
\def\eieobb{\eta_{\bbar}^{E,0}}

\def\eq#1{eq.~(\ref{#1})}
\def\eqs#1#2{eqs.~(\ref{#1}) and (\ref{#2})}

\def\eqst#1#2{eqs.~(\ref{#1})--(\ref{#2})}
\def\eqthree#1#2#3{eqs.~(\ref{#1}), (\ref{#2}) and (\ref{#3})}
\def\Eq#1{Eq.~(\ref{#1})}

\def\ifmath#1{\relax\ifmmode #1\else $#1$\fi}
\def\ls#1{\ifmath{_{\lower1.5pt\hbox{$\scriptstyle #1$}}}}
\def\lss#1{\ifmath{^{\,\lower2.5pt\hbox{$\scriptstyle #1$}}}}
\def\lsup#1{^{\lower 6pt\hbox{$\scriptstyle#1$}}}
\def\llsup#1{^{\lower 3pt\hbox{$\scriptstyle#1$}}}
\def\lasup#1{^{\lower 2pt\hbox{$\scriptstyle#1$}}}

\def\half{\tfrac{1}{2}}

\def\lsup#1{^{\lower 6pt\hbox{$\scriptstyle#1$}}}

\def\ddel{\!\!\mathrel{\raise1.5ex\hbox{$\leftrightarrow$\kern-.85em
\lower1.7ex\hbox{$\partial$}}}}

\def\phab{\phantom{ab}}
\newcommand{\be}{\begin{equation}}
\newcommand{\ee}{\end{equation}}
\newcommand{\ba}{\begin{align}}
\newcommand{\ea}{\end{align}}
\newcommand{\bea}{\begin{eqnarray}}
\newcommand{\eea}{\end{eqnarray}}

\def\lsim{\mathrel{\rlap{\lower4pt\hbox{\hskip1pt$\sim$}}
    \raise1pt\hbox{$<$}}}         
\def\gsim{\mathrel{\rlap{\lower4pt\hbox{\hskip1pt$\sim$}}
    \raise1pt\hbox{$>$}}}         

\begin{document}

\title{
\vspace*{-3.3cm}
\begin{flushright}
\normalsize{
SCIPP-15/07
}
\end{flushright}
\vspace{0.5cm}
\Large
\textbf{Preserving the validity of the Two-Higgs Doublet Model up to the Planck scale}
\\ 
}\vspace*{2.0cm}

\author{P.M.~Ferreira}
\email{ferreira@cii.fc.ul.pt}
\affiliation{ISEL - Instituto Superior de Engenharia de Lisboa, Instituto Polit\'ecnico de Lisboa, Portugal}
\affiliation{Centro de F\'{i}sica Te\'{o}rica e Computacional, Faculdade de Ci\^{e}ncias, Universidade de Lisboa, Portugal}
\author{Howard E.~Haber}
\email{haber@scipp.ucsc.edu}
\author{Edward Santos}
\email{edrsanto@ucsc.edu}
\affiliation{Santa Cruz Institute for Particle Physics, University of California, Santa Cruz, California 95064, USA}

\begin{abstract}
 We examine the constraints on the two Higgs doublet model (2HDM) due
 to the stability of the scalar potential and absence of Landau poles
 at energy scales below the Planck scale.  We employ the most general
 2HDM that incorporates an approximately Standard Model (SM) Higgs boson
 with a flavor aligned Yukawa sector to eliminate potential
 tree-level Higgs-mediated flavor changing neutral currents.
 Using basis independent techniques, we exhibit regimes
 of the 2HDM parameter space with a 125 GeV SM-like Higgs boson
 that is stable and perturbative up to the Planck scale.  Implications for the heavy
 scalar spectrum are exhibited.

\end{abstract}

\maketitle

%
%
%
%

\section{\label{sec:intro}Introduction}

With the recent discovery of a Higgs-like particle with a mass of about 125 GeV by both the
ATLAS \cite{:2012gk} and CMS \cite{:2012gu} collaborations, the focus has now turned to
deciphering the properties of this particle and determining whether it is the Standard Model
(SM) Higgs particle, or part of an extended Higgs sector.
Analyses performed by ATLAS and CMS collaborations have shown
that the couplings of the newly discovered particle are
consistent with a SM-like Higgs boson, within the accuracy of their
measurements.  In light of the present precision of the Higgs data,
the LHC can claim to have discovered a SM-like Higgs boson.
However, there is still plenty of room for deviations from SM behavior
of $\mathcal{O}(10\%)$.  A SM-like Higgs boson is easily
achieved in an extended Higgs sector in the decoupling limit, where
the lightest scalar is identified as the observed SM-like Higgs boson,
and the heavier scalars are somewhat separated in mass (e.g.~with a
mass scale above 350 GeV~\cite{Maiani:2013nga,Arbey:2013jla}).

Before the mass of the Higgs boson was known, upper bounds on the SM
Higgs mass were obtained by requiring that the running quartic
coupling parameter avoid Landau poles (LPs), i.e., the coupling was
required to remain finite up to a given energy scale $\Lambda$~\cite{Cabibbo:1979ay,Lindner:1985uk,Hambye:1996wb}. Lower
bounds were obtained by requiring that the scalar potential remain stable
during renormalization group (RG) evolution~\cite{Sher:1988mj,Lindner:1988ww,Ford:1992mv,Sher:1993mf,Altarelli:1994rb,Casas:1994qy}. That is, the scalar potential
is bounded from below
at all scales between the electroweak scale and $\Lambda$.
These bounds were contingent on the assumption that no new physics
beyond the SM (BSM) enters between the electroweak scale and $\Lambda$.
Turning around the argument, the existence of a LP or an instability
of the scalar potential at some energy scale $\Lambda$ suggests that
new BSM physics must be present at or below $\Lambda$.

After the discovery of the Higgs boson,
previously obtained bounds were updated using two-loop renormalization group equations
(RGEs) in Ref.~\cite{Degrassi:2012ry} and three-loop RGEs by
Ref.~\cite{Bezrukov:2012sa}.  The most recent analysis of Ref.~\cite{Buttazzo:2013uya}
has shown that the SM scalar potential
becomes unstable at a value of $\Lambda$ well below the Planck scale, if the Higgs boson mass is smaller
than $129.6\pm 1.5$~GeV.\footnote{The quoted uncertainty takes into
 account the parametric uncertainty of $m_t$ and $\alpha_s$ and the
 effects of unknown higher order corrections~\cite{Buttazzo:2013uya}.}
Taken at face value, these results would further imply that we live in a metastable vacuum that will eventually
(and catastrophically) decay via tunneling into the true vacuum.   However, the lifetime of the
metastable vacuum is many orders of magnitude larger than the age of the universe~\cite{Buttazzo:2013uya,Espinosa:1995se}.
On the other hand, if the electroweak vacuum is absolutely stable, then
the recent LHC discovery of a 125 GeV SM-like Higgs boson requires the
existence of new BSM physics at an energy scale below a scale of
$\Lambda\simeq 10^{9.5}$~GeV, where there is an uncertainty of about
1 in the exponent due to parametric uncertainties of $m_t$, $\alpha_s$
and the Higgs mass~\cite{Buttazzo:2013uya}, in order to avoid the metastability of the SM vacuum.

Although the prospect of existence of new BSM physics is exciting,
there is no guarantee that the scale of the new physics is close to
the electroweak scale.  Nevertheless, arguments motivated by naturalness
of the electroweak symmetry breaking (EWSB) mechanism suggest that BSM physics
should be present at or near the TeV scale (see e.g., Refs.~\cite{Giudice:2008bi,Wells:2013tta}
for a review and a guide to the literature).  Many models of
new physics have been proposed to address the origin of ESWB, and
many of these approaches possess extended Higgs sectors.
However, in such models one must specify the BSM physics in order to
study the behavior of running couplings between the electroweak scale
and some very high energy scale $\Lambda$.  At present, there is no
direct experimental evidence that the origin of the EWSB scale is a consequence of naturalness.  Adding additional Higgs multiplets at or near the TeV
scale by themselves does not address the origin of EWSB.  Indeed, one
could argue that it makes matters worse by adding additional
fine-tuning constraints.   Nevertheless, in this paper we shall accept
the fine-tunings required to sustain an extended Higgs sector near the
TeV scale.  After all, we know that multiple generations exists in the
fermionic sector of the Standard Model.   Thus, we should be prepared for the possibility that
the scalar sector of the theory is also non-minimal.

Here, we shall focus on the two-Higgs doublet model (2HDM),
which was initially proposed by Lee in 1973~\cite{Lee:1973iz} (for a
review, see e.g.~Ref.~\cite{Branco:2011iw}). It provides a richer
Higgs particle spectrum, namely three neutral scalars and a charged
pair.  The 2HDM admits the possibility of CP-violation in the scalar
potential, both explicit or spontaneous. In the limit of
CP-conservation, two of the neutral scalars are CP-even, typically
denoted by $h$ and $H$, (where $m_h<m_H$) while the other neutral
scalar is CP-odd, denoted by $A$.  We shall consider a very general
version of the 2HDM that is not inconsistent with present data.  Such
a model must possess a SM-like Higgs boson (within the accuracy of the
present Higgs data).  In addition, Higgs-mediated tree-level flavor
changing neutral currents (FCNCs) must be either absent or highly
suppressed.  These conditions are achieved if the non-minimal Higgs
states of the model have masses above about 350 GeV and if the Yukawa
couplings are aligned in such a way that the neutral Higgs couplings
are diagonal in the mass-basis for the neutral Higgs bosons.  The most
general 2HDM parameter space allowed by the present data is somewhat larger
than the one specified here.  Nevertheless, the restricted parameter
space outlined above is still quite general and incorporates the more
constrained 2HDMs considered in the literature.

The existence of additional scalar degrees of freedom in an extended
Higgs sector provides an opportunity to cure the vacuum metastability
problem of the SM Higgs boson.  However, by demanding no Landau poles
and requiring a stable scalar potential at all energy scales up to the
Planck scale, one imposes strong constraints on the parameter space of
the extended Higgs sector.  Investigations of this type have been
performed in extended Higgs sectors prior to the discovery of the
Higgs boson in
Refs.~\cite{Kreyerhoff:1989fa,Freund:1992yd,Kastening:1992by,Nie:1998yn,Kanemura:1999xf,Ferreira:2009jb,Gonderinger:2009jp}.
With the discovery and identification of a SM-like Higgs boson, 
the question of the validity of extended Higgs sectors up to the
Planck scale has become more focused.  A number of authors have considered
the stability properties of extended Higgs sectors with additional singlet scalar
fields~\cite{EliasMiro:2012ay,Lebedev:2012zw,Pruna:2013bma,Costa:2014qga}
and 2HDMs with constrained scalar potentials~\cite{Chakrabarty:2014aya,Das:2015mwa,Chowdhury:2015yja}.


In this paper, we examine the theoretical consistency
of \textit{the most general}
2HDM between the electroweak scale and the Planck scale, using the
one-loop RGEs of the model to investigate the possible occurrence of Landau
poles and instability of the scalar potential.  We focus on the
decoupling regime of the 2HDM where the 125 GeV Higgs boson is SM-like~\cite{Gunion:2002zf,Haber:2013mia},
and assume Yukawa alignment in the flavor sector~\cite{Pich:2009sp} to avoid
Higgs-mediated tree-level FCNCs.  Our aim is to exhibit
the allowed regions of the 2HDM parameter space that are free from both
Landau poles and vacuum instability below the
Planck scale.  In particular, a 2HDM that satisfies these constraints
does not require further BSM physics to stabilize the theory.

One of the distinguishing features of the most general 2HDM is the fact that the two scalar doublet, hypercharge-one fields are indistinguishable.   One is always free to define new linear combinations of the scalar doublets that preserve
the kinetic energy terms of the Lagrangian.  A specific choice for the scalar fields is called a \textit{basis},
and any physical prediction of the theory must be \textit{basis independent}.
In our analysis, we employ a basis-independent formalism introduced in Ref. \cite{Davidson:2005cw}.
We consider the most general 2HDM scalar potential (which is potentially CP-violating) and the most general Yukawa sector, which introduces three additional independent $3\times 3$ matrix Yukawa couplings.
Without additional assumptions, the latter yields Higgs-mediated tree-level FCNCs, in conflict with observed data.
In order to circumvent this, we
impose a ``flavor alignment ansatz", introduced in Ref. \cite{Pich:2009sp}, which postulates that the
independent matrix Yukawa couplings are proportional to the corresponding quark and charged lepton mass matrices.
In this case one finds that, in the mass basis for the quarks and leptons, the matrix Yukawa couplings are flavor diagonal,
and the Higgs-mediated tree-level FCNCs are absent.   One way to achieve alignment in the Yukawa sector is to introduce a set of discrete symmetries which constrain the Higgs scalar potential and Yukawa couplings.   The so-called Type-I and II
2HDMs~\cite{Hall:1981bc},
and the related Type X and Type Y 2HDMs~\cite{Barger:1989fj,Aoki:2009ha}
provide examples of this type.  Indeed, Ref.~\cite{Ferreira:2010xe}
showed that the flavor alignment is preserved under RGE running if and
only if such discrete symmetries are present.  The flavor alignment ansatz is more general, but requires fine-tuning in the absence of an underlying symmetry.

This paper is organized as follows: In section \ref{sec:2HDM}, we
review the basis-independent formalism as applied to the 2HDM. In section
\ref{sec:Yukawa}, we describe the Yukawa sector and present the flavor
alignment model used in this analysis. In section \ref{sec:Results},
we present our numerical analysis of the mass bounds governing the
lightest scalar, which are derived by requiring the stability of the
2HDM potential and the absence of Landau poles in the scalar quartic
couplings below the Planck scale.  Our analysis employs both the one-loop
RG running of the quartic couplings, along with an
estimate of the effects of the two-loop corrections.
In section \ref{sec:Conclusions}, we present our conclusions. The one-loop basis independent
RGEs are presented in Appendix \ref{sec:RGEs}, and stability conditions on the basis-independent
2HDM potential are derived in Appendix~\ref{sec:Stability}.

%
%
%
%

\section{Basis-Independent Treatment of the 2HDM}
\label{sec:2HDM}

\subsection{\label{sec:hbasis}The Higgs Basis}

In a generic basis, the most general renormalizable
SU(3)$_C\times$SU(2)$_L\times$U(1)$_Y$ gauge-invariant
2HDM scalar potential is given by
\beqa
{\mathcal{V}} &=& m_{11}^2\bigl(\Phi^{\dagger}_1\Phi_1\bigr) + m_{22}^2\bigl(\Phi^{\dagger}_2\Phi_2\bigr) -
\bigl[m_{12}^2\Phi^{\dagger}_1\Phi_2 +{\rm h.c.}\bigr] \nonumber \\[8pt]
&&+\tfrac{1}{2}\lambda_1\bigl(\Phi^{\dagger}_1\Phi_1\bigr)^2+\tfrac{1}{2}\lambda_2\bigl(\Phi^{\dagger}_2\Phi_2\bigr)^2+\lambda_3\bigl(\Phi^{\dagger}_1\Phi_1\bigr)\bigl(\Phi^{\dagger}_2\Phi_2\bigr) + \lambda_4\bigl(\Phi^{\dagger}_1\Phi_2\bigr)\bigl(\Phi^{\dagger}_2\Phi_1\bigr)
\nonumber \\[8pt]
&&+ \{\tfrac{1}{2}\lambda_5\bigl(\Phi^{\dagger}_1\Phi_2\bigr)^2 +\bigl[\lambda_6\bigl(\Phi^{\dagger}_1\Phi_1\bigr)+\lambda_7\bigl(\Phi^{\dagger}_2\Phi_2\bigr)\bigr]\bigl(\Phi^{\dagger}_1\Phi_2\bigr) + {\rm h.c.}\bigr\} \,, \label{eq:generic}
\eeqa

\noindent where $\Phi_1, \Phi_2$ are two hypercharge-one complex
scalar doublets. The two doublets separately acquire vacuum
expectation values (vevs) $\langle \Phi^0_1 \rangle = v_1 / \sqrt{2}$
and $\langle \Phi^0_2 \rangle = v_2/ \sqrt{2}$ with the constraint
$v^2 = |v_1|^2+|v_2|^2 \simeq (246 \; \text{GeV})^2$. The parameters
$\lambda_{1,2,3,4}$ and $m_{11}^2, m_{22}^2$ are real whereas
$\lambda_{5,6,7}$ and $m_{12}^2$ are potentially complex. The 2HDM
is CP-conserving if there exists a basis in which all of the
parameters and the vacuum expectation values are simultaneously
real.

We shall adopt a basis-independent formalism as developed in
Ref.~\cite{Davidson:2005cw}, which provides basis-independent 2HDM
potential parameters that are invariant under a global U(2)
transformation of the two scalar doublet fields, $\Phi_a\to U_{a\bar{b}}\Phi_b$ ($a$, $\bar{b}=1,2$).

It is convenient to define the so-called \textit{Higgs basis} of scalar doublet fields,
\begin{equation} \label{Hbasis}
H_1=\begin{pmatrix}H_1^+\\ H_1^0\end{pmatrix}\equiv \frac{v^*_1 \Phi_1+v^*_2\Phi_2}{v}\,,
\qquad\quad H_2=\begin{pmatrix} H_2^+\\ H_2^0\end{pmatrix}\equiv\frac{-v_2 \Phi_1+v_1\Phi_2}{v}
 \,,
\end{equation}
so that $\langle{H_1^0}\rangle=v/\sqrt{2}$ and $\langle{H_2^0}\rangle=0$.
The Higgs basis is uniquely defined up to a rephasing of the $H_2$ field, $H_2\to e^{i\chi}H_2$.
In the Higgs basis, the scalar potential takes the familiar form,\footnote{As discussed in Appendix A, the squared-mass and coupling coefficients,
$Y_1, Y_2,$ and $Z_{1,2,3,4}$ can be expressed as U(2)-invariant combinations of the scalar potential coefficients and the vevs, whereas $Y_3$ and $Z_{5,6,7}$  are U(2)-pseudoinvariant combinations of the scalar potential coefficients and the vevs that are rephased under a U(2) transformation~\cite{Davidson:2005cw}.}
\beqa\label{eq:HiggsBasis}
{\mathcal{V}} &=& Y_1\bigl(H^{\dagger}_1H_1\bigr) + Y_2\bigl(H^{\dagger}_2H_2\bigr) + \bigl[Y_3H^{\dagger}_1H_2 +{\rm h.c.}\bigr] +\tfrac{1}{2}Z_1\bigl(H^{\dagger}_1H_1\bigr)^2+\tfrac{1}{2}Z_2\bigl(H^{\dagger}_2H_2\bigr)^2 \nonumber \\[8pt]
&&+Z_3\bigl(H^{\dagger}_1H_1\bigr)\bigl(H^{\dagger}_2H_2\bigr) + Z_4\bigl(H^{\dagger}_1H_2\bigr)\bigl(H^{\dagger}_2H_1\bigr) +
\{\tfrac{1}{2}Z_5\bigl(H^{\dagger}_1H_2\bigr)^2 + \bigl[Z_6\bigl(H^{\dagger}_1H_1\bigr)  \nonumber \\[8pt]
&&+Z_7\bigl(H^{\dagger}_2H_2\bigr)\bigr]\bigl(H^{\dagger}_1H_2\bigr) +
       {\rm h.c.}\bigr\} \,,
\eeqa
where $Y_1, Y_2,$ and $Z_{1,2,3,4}$ are real parameters and
uniquely defined, whereas $Y_3$ and $Z_{5,6,7}$  transform under a rephasing of $H_2$, viz.,
$[Y_3,Z_6,Z_7]\to e^{-i\chi}[Y_3,z_6,Z_7]$ and $Z_5\to e^{-2i\chi}Z_5$. Minimizing the scalar potential then yields
\begin{equation}
Y_1 = -\tfrac{1}{2} Z_1 v^2, \;\;\;\; Y_3 = -\tfrac{1}{2} Z_6 v^2.
\end{equation}
The scalar potential is CP-violating if no choice of $\chi$ can be found in which all Higgs basis scalar potential parameters are simultaneously real.

The tree-level mass eigenstates of the neutral scalars  can be obtained by diagonalizing the neutral scalar squared-mass matrix in the Higgs basis~\cite{Branco:1999fs,Haber:2006ue},
\begin{equation}
{\mathcal{M}} = v^2
\begin{pmatrix}
Z_1 & \text{Re}(Z_6) & -\text{Im}(Z_6) \\
\text{Re}(Z_6) & \frac{1}{2}[Z_3+Z_4+\text{Re}(Z_5)]+Y_2/v^2 &  -\frac{1}{2}\text{Im}(Z_5)  \\
-\text{Im}(Z_6) & -\frac{1}{2}\text{Im}(Z_5) & \frac{1}{2}[Z_3+Z_4-\text{Re}(Z_5)]+Y_2/v^2 \\
\end{pmatrix}.
\end{equation}

The diagonalizing matrix is a real orthogonal $3\times 3$ matrix that is parameterized by three mixing angles
$\theta_{12}, \theta_{13},$ and $\theta_{23}$ (details can be found in Ref.~\cite{Haber:2006ue}).
In terms of U(2)-invariant combinations of the mixing angles and
scalar potential parameters, the squared-masses of the three neutral
Higgs bosons, denoted by $h_1$, $h_2$ and $h_3$ respectively, are given by~\cite{Haber:2006ue},
\beqa
m_k^2 &=& |q_{k2}|^2\ Y_2+ v^2\biggl\{q_{k1}^2 Z_1 +  \half |q_{k2}|^2\bigl[Z_3+Z_4-\text{Re}(Z_5 e^{-2i\theta_{23}})\bigr]
\nonumber \\[6pt]
&&\qquad\qquad\qquad +\text{Re}(q_{k2})\text{Re}(q_{k2} Z_5 e^{-2i\theta_{23}}) + 2q_{k1}\text{Re}(q_{k2}Z_6 e^{-i\theta_{23}})\biggr\},\quad \text{for $k=1,2,3$},\label{eq:masses}
\eeqa
where the $q_{ki}$ are invariant combinations of the mixing angles shown in Table \ref{table:qs}.
It is convenient to choose a convention where $m_1<m_2<m_3$ (which can
always be arranged by an appropriate choice of neutral Higgs mixing angles).
The squared-mass of the charged scalars is given by
\begin{equation} \label{mchhiggs}
m_{H^\pm}^2 = Y_2 + \tfrac{1}{2}Z_3v^2.
\end{equation}

\begin{table}
\caption{$q_{ki}$ as a function of the neutral Higgs mixing angles in the Higgs basis.}
\begin{tabular}{|c|| c | c |}
\hline
$k$ & $q_{k1}$ & $q_{k2}$ \\
\hline
$\phab 1\phab$ & \phab $\cos\theta_{12}\cos\theta_{13}$\phab  &\phab  $-\sin\theta_{12} -i\cos\theta_{12}\sin\theta_{13}$ \phab \\
$\phab 2\phab $ & \phab $\sin\theta_{12}\cos\theta_{13}$\phab  & \phab $\cos\theta_{12} -i\sin\theta_{12}\sin\theta_{13}$\phab  \\
$\phab 3\phab $ &\phab  $\sin\theta_{13}$\phab  & \phab $i\cos\theta_{13}$\phab  \\
\hline
\end{tabular}
\label{table:qs}
\end{table}

\subsection{\label{sec:decouple}Decoupling Limit}

The decoupling limit corresponds to taking the squared-mass parameter
of the Higgs basis field $H_2$ large while holding the Higgs
quartic coupling parameters fixed.  In the perturbative regime, we take
$|Z_i|\lsim\mathcal{O}(1)$ and $Y_2\gg v^2$.  In this case~\cite{Haber:2006ue,Haber:2013mia},
\beq
\sin\theta_{12}\sim\sin\theta_{13}\sim\mathcal{O}\left(\frac{v^2}{Y_2}\right)\,.
\eeq
In addition, the decoupling limit requires that
\beq \label{imz5}
{\rm Im}(Z_5 e^{-2i\theta_{23}})\sim\mathcal{O}\left(\frac{v^2}{Y_2}\right)\,,
\eeq
which implies that
\beq \label{signamb}
{\rm Re}(Z_5 e^{-2i\theta_{23}})=-|Z_5|\,.
\eeq
The overall sign in \eq{signamb} [which is not determined by
  \eq{imz5}]
is fixed in the convention where $m_2<m_3$.
Using the above results in \eq{eq:masses} yields
\beqa
m_1^2&=&Z_1
v^2\left[1+\mathcal{O}\left(\frac{v^2}{Y_2}\right)\right]\,,
\label{eqn:m1}\\
m_2^2&=& Y_2+\half v^2\left[Z_3+Z_4-|Z_5|+
\mathcal{O}\left(\frac{v^2}{Y_2}\right)\right]\,,\label{eqn:m2} \\
m_3^2&=& Y_2+\half v^2\left[Z_3+Z_4+|Z_5|+
\mathcal{O}\left(\frac{v^2}{Y_2}\right)\right]\,.\label{eqn:m3}
  \eeqa
At energy scales below $Y_2$, the effective low-energy theory
corresponds to the Standard Model with one Higgs doublet.
Consequently, in the decoupling limit
the properties of $h_1$ approach those of the SM Higgs boson.
The non-minimal Higgs states are roughly degenerate in mass,
$m^2_2\sim m^2_3\sim m^2_{H^\pm}\sim Y_2$, with squared-mass splittings
of $\mathcal{O}(v^2)$,
\beqa
m_3^2-m_2^2&\simeq &|Z_5|v^2\,,\label{eqn:masssplit1}\\
m_3^2-m_{H^\pm}^2&\simeq &\half\bigl(Z_4+|Z_5|\bigr)v^2\,.\label{eqn:masssplit2}
\eeqa

In the decoupling limit of a general 2HDM, the tree-level CP-violating
and flavor-changing neutral Higgs couplings of the SM-like Higgs state
$h_1$ are suppressed by factors of ${\mathcal{O}}(v^2/Y_2^2)$. The
corresponding interactions of the heavy neutral Higgs bosons ($h_2$
and $h_3$) and the charged Higgs bosons ($H^\pm$) can exhibit both
CP-violating and flavor non-diagonal couplings.  If $Y_2$ is
sufficiently large, then FCNCs mediated by the lightest neutral scalar
can be small enough to be consistent with experimental data.  However,
for values of $Y_2$ of order 1~TeV and below, tree-level
Higgs-mediated FCNCs are problematical in the case of a generic Yukawa
sector.

%
%
%
%

\section{Yukawa Sector}\label{sec:Yukawa}

The most general 2HDM Yukawa sector, describing Higgs-fermion interactions, includes six Yukawa matrices (as compared to three in the SM). In a generic basis, the Yukawa Lagrangian for the Higgs--quark interactions is given by \eq{ymodeliii0}.
Following the discussion of Appendix~A, we can re-express the Yukawa Lagrangian in terms of the quark mass-eigenstate fields~\cite{Haber:2010bw},
\beqa
-{\mathcal{L}}_Y &=& \overline{U}_L \bigl(\eta_1^U\Phi^{0*}_1+\eta_2^U\Phi^{0*}_2\bigr)- \overline{D}_L K^{\dagger}
\bigl(\eta^U_1\Phi^{-}_{1}  +\eta^U_2\Phi^{-}_{2} \bigr)U^R \nonumber \\[6pt]
 && + \overline{U}_L K \bigl(\eta^{D\dagger}_{1}\Phi^{+}_{1} +\eta^{D\dagger}_{2}\Phi^{+}_{2}\bigr)  D^R +
 \overline{D}_L \bigl(\eta^{D\dagger}_{1}\Phi^{0}_{1}+ \eta^{D\dagger}_{2}\Phi^{0}_{2}\bigr) D^R + \text{h.c.},\label{lyukgeneric}
\eeqa
where $\eta^{U,D}_{1,2}$ are $3\times3$ Yukawa coupling matrices and $K$ is the CKM matrix.

Using \eq{Hbasis}, one can rewrite \eq{lyukgeneric} in terms of the Higgs basis scalar doublet fields,
\beqa
-{\mathcal{L}}_Y &=& \overline{U}_L (\kappa^U H^{0\dagger}_1 + \rho^U H^{0\dagger}_2) U^R - \overline{D}_L K^\dagger (\kappa^U H^{-}_1 + \rho^U H^{-}_2) U^R  \nonumber \\[6pt]
 & &  +\overline{U}_L K (\kappa^{D\dagger} H^{+}_1
+ \rho^{D\dagger} H^{+}_2) D^R+ \overline{D}_L (\kappa^{D\dagger} H^{0}_1 + \rho^{D\dagger} H^{0}_2) D^R + \text{h.c.},
\label{hbasisyuk}
\eeqa
where\footnote{As noted in \eq{rhotrans}, the $\rho^Q$ are U(2)-pseudoinvariant combinations of the Yukawa coupling matrices and the vevs, whereas the $\kappa^Q$ are U(2)-invariants.}
\beq \label{kr}
\kappa^Q\equiv \frac{v_1^*\eta_1^Q+v_2^*\eta_2^Q}{v}\,,\qquad\quad
\rho^Q\equiv \frac{-v_2\eta_1^Q+v_1\eta_2^Q}{v}\,.
\eeq
Note that $\rho^Q\to e^{-i\chi}\rho^Q$ with respect to the rephasing $H_2\to e^{i\chi}H_2$.
Since $\langle H_1^0 \rangle = v/\sqrt{2}$ and $\langle H_2^0 \rangle = 0$, it follows that the
$\kappa^{U,D}$ are proportional to the diagonal quark mass matrices, $M_U$ and $M_D$, whose matrix elements are real and non-negative,
\begin{equation} \label{Mud}
M_U = \frac{v\kappa^U}{\sqrt{2}}  = \text{diag}(m_u,m_c,m_t), \qquad\quad M_D = \frac{v\kappa^{D}}{\sqrt{2}}=\text{diag}(m_d,m_s,m_b)\,.
\end{equation}
The Yukawa couplings of the Higgs doublets to the leptons can be similarly treated by replacing $U\to N$, $D\to E$, $M_U\to 0$, $M_D\to M_E$ and $K\to \mathds{1}$, where $N=(\mu_e,\nu_\mu,\nu_\tau)$, $E=(e,\mu,\tau)$ and $M_E$ is the diagonal charged lepton mass matrix.

Since the Yukawa matrices $\rho^{U,D,E}$ are independent complex $3\times3$ matrices, it follows that the Yukawa Lagrangian exhibited in \eq{hbasisyuk} generically exhibits tree-level Higgs mediated FCNCs.
The off-diagonal elements of the $\rho^{U,D}$ matrices are highly constrained by experimental data to be very small.   As first shown by Glashow, Weinberg and Paschos (GWP)~\cite{Glashow:1976nt,Paschos:1976ay}, it is possible to \textit{naturally} eliminate tree-level Higgs mediated FCNCs if, for some choice of basis of the scalar fields,
at most one Higgs multiplet is responsible for providing mass for quarks or leptons of a given electric
charge.  In the 2HDM, the GWP condition is usually imposed in four different ways by employing the appropriate
$\mathbb{Z}_2$ discrete symmetry~\cite{Haber:1978jt,Donoghue:1978cj,Hall:1981bc,Barger:1989fj,Aoki:2009ha}:
\begin{enumerate}
\item Type-I Yukawa couplings: $\eta_1^U=\eta_1^D=\eta_1^L=0$,
\item Type-II Yukawa couplings: $\eta_1^U=\eta_2^D=\eta_2^L=0$.
\item Type-X Yukawa couplings: $\eta_1^U=\eta_1^D=\eta_2^L=0$,
\item Type-Y Yukawa couplings: $\eta_1^U=\eta_2^D=\eta_1^L=0$.
\end{enumerate}
For example, it follows from \eq{kr} that in the Type-I 2HDM,
\beq \label{t1}
\rho^{U,D,L}=\frac{v_1}{v_2^*}\,\kappa^{U,D,L}\,,
\eeq
and in the Type-II 2HDM,
\beq \label{t2}
\rho^U=\frac{v_1}{v_2^*}\,\kappa^U\,,\qquad \rho^{D,L}=-\frac{v_2}{v_1^*}\,\kappa^{D,L}\,.
\eeq
In light of \eq{Mud}, the $\rho^F$ ($F=U,D,L$) are, in these cases, diagonal matrices in which case the neutral Higgs--fermion Yukawa interactions are flavor-diagonal at tree-level.

If only phenomenological considerations are invoked in choosing the Higgs--fermion Yukawa couplings, then it is possible to consider the more general case of the flavor-aligned 2HDM introduced in Ref. \cite{Pich:2009sp}.
In this model applied to the Higgs basis, one imposes the following conditions
\begin{equation}\label{eqn:align}
\rho^U = \alpha^U \kappa^U, \;\;\;\; \rho^{D} = \alpha^{D} \kappa^{D}, \;\; \text{and} \;\; \rho^L = \alpha^L \kappa^L,
\end{equation}
which generalize the Type-I and II results exhibited in \eqs{t1}{t2}.     In \eq{eqn:align},
the alignment parameters, $\alpha^{U,D,L}$, are arbitrary complex constants.\footnote{In practice, if the magnitude of the alignment constants are too large, then some of the Higgs-fermion Yukawa couplings will develop Landau poles below the Planck scale.  In our analysis, we will determine the allowed regions of the flavor-aligned  2HDM parameter space where such Landau poles are absent.}
The flavor alignment condition shown in \eq{eqn:align} is not imposed by any symmetry, and is strictly unnatural (i.e., it can be achieved only by a fine-tuning of the model parameters).
Equivalently, as observed in Ref.~\cite{Ferreira:2010xe}, the flavor alignment is preserved under RGE running only in the case of Type I, II, X and~Y Yukawa couplings.  Nevertheless, one can imagine the possibility of new dynamics above the electroweak scale that could be responsible for an approximately flavor-aligned 2HDM.   Thus, in our analysis we shall employ the more general
\eq{eqn:align}, which is sufficient for satisfying the phenomenological FCNC constraints.\footnote{By choosing to work in the decoupling limit where $Y_2\gsim 500$~GeV, we ensure that FCNCs generated by one-loop radiative effects are not too large
to be in conflict with experimental data (see e.g.~Ref.~\cite{Mahmoudi:2009zx}).}

%
%
%
%

\section{\label{sec:Results}RG stability and perturbativity of the 2HDM}

Let us assume that the observed SM-like Higgs boson (with $m_h\simeq 125$~GeV)
is part of a 2HDM in the decoupling limit with a flavor-aligned Yukawa sector, with no other new physics present
beyond the 2HDM below the Planck scale.\footnote{Incorporating light neutrino masses via
the seesaw mechanism~\cite{seesaw} with the mass scale of the right-handed neutrino sector assumed to be of order a typical grand unified scale has a very minor impact on the considerations in this paper.}
We shall examine whether there are regions of the 2HDM parameter space that yield a consistent model under RG running from the electroweak to the Planck scale.   In general, two potential problems can arise
in the RG evolution.  First,  Landau poles could arise from the divergence of the 2HDM quartic scalar couplings and/or Yukawa couplings.  Second, the 2HDM scalar potential could become unstable at a higher energy scale. The case of Landau poles is fairly straightforward, although the precise energy scale at which they arise cannot be strictly determined, since it lies outside the perturbative regime of the RGEs.  In practice, we shall consider that a Landau pole occurs when the relevant coupling exceeds 100 for some energy scale $\Lambda \leq M_{Pl}$.  Indeed, once such a large coupling is reached, it will very quickly diverge
at an energy scale very close to $\Lambda$.
In our analysis, we employ the one-loop RGEs for the quartic scalar couplings of the 2HDM in the Higgs basis given in Appendix \ref{sec:RGEs}.  These equations are strongly coupled,
and thus a divergence in one quartic scalar coupling will cause a divergence in the rest.  The leading effects of two-loop running will be assessed at the end of this section.

In the SM, the requirement that the scalar potential is stable at all energy scales below the scale $\Lambda$ is easily implemented.  It is sufficient to require that the SM quartic scalar coupling is positive, i.e.~$\lambda_{\rm SM} (\Lambda) > 0$ for
$\Lambda>v$.
Requiring that the 2HDM scalar potential is stable at all energy scales below the scale $\Lambda$ leads to a more complicated
set of conditions.  In the 2HDM with an unbroken, or softly broken, $\mathbb{Z}_2$ discrete symmetry that sets $\lambda_6=\lambda_7=0$ in \eq{eq:generic}, the stability conditions were first obtained in Ref.~\cite{Deshpande:1977rw},
\beqa \label{eq:cpstable}
\lambda_1 &>& 0 \,, \label{eq:cpstable1} \\
\lambda_2 &>& 0 \,, \label{eq:cpstable2} \\
\lambda_3 &>& -\sqrt{\lambda_1 \lambda_2} \,, \label{eq:cpstable3} \\
\lambda_3 + \lambda_4 - |\lambda_5| &>& - \sqrt{\lambda_1 \lambda_2} \,. \label{eq:cpstable4}
\eeqa
However, in the case of a completely general scalar potential, the corresponding stability conditions are far more complicated (with no simple analytic form). Ref.~\cite{Ivanov:2006yq} provides an algorithm for deriving the stability conditions for a general 2HDM, with no symmetry or CP assumptions imposed on the 2HDM scalar potential.  In terms of the Higgs basis parameters, this algorithm is summarized in Appendix \ref{sec:Stability}.  Except for special cases for the quartic scalar couplings, the corresponding stability conditions must be determined numerically.

We now describe in detail the procedure used in our analysis.  We assume that we are in the decoupling regime of the 2HDM, where the mass scale of the heavy Higgs sector is of $\mathcal{O}(\Lambda_H)$.  In light of \eqthree{mchhiggs}{eqn:m2}{eqn:m3}, we henceforth set $\Lambda^2_H\equiv Y_2$.
\begin{enumerate}
\item Start with the SM Higgs potential defined at the scale of the 125 GeV Higgs boson.
\item Use SM RG evolution to run the Higgs-self coupling parameter $\lambda$ and the fermion mass matrices up to the scale $\Lambda_H$.\footnote{Starting the RG evolution at $m_Z$, we use a five flavor scheme to run up to $m_t$ and a six flavor scheme above $m_t$.  Running quark mass masses at $m_Z$ and $m_t$ are obtained from the
RunDec Mathematica software package~\cite{Chetyrkin:2000yt}, based on quark masses provided in Ref.~\cite{pdg}.  For simplicity, the effects of the lepton masses are ignored, as these contribute very little to the running of the $Z_i$.}
\item Match the one-doublet Higgs potential with the 2HDM potential by taking $Z_1 = \lambda(\Lambda_H)$
and $\kappa^F=\sqrt{2}M_F(\Lambda_H)/v$ (for $F=U,D$).  This establishes the low energy boundary conditions.
The effects of the lepton masses are negligible and have been ignored.
\item Scan over all other 2HDM quartic scalar coupling parameters $Z_i$ and Yukawa alignment parameters $\alpha^F$ ($F=U,D$).
The latter fix the values of the $\rho^F(\Lambda_H)$.
\item Run the 2HDM RGEs for the $Z_i$, $\kappa^F$ and $\rho^F$
up to higher energies $\Lambda$. Check for stability of the potential at the scale $\Lambda$ using the procedure summarized in Appendix \ref{sec:Stability}.
\item Stop the running if a Landau pole is encountered or if the stability conditions cannot be satisfied.
\end{enumerate}

For the scalar sector, we scanned over the parameter space using 100,000 points, with $|Z_i | \lesssim {\mathcal{O}}(1)$,  for $i =2,...,7$, to enforce the decoupling limit. These points were also subject to the constraint that they obey the stability conditions presented in Appendix \ref{sec:Stability}. Note that when $|Z_i | \ll 1$ for $i=2,...,7$, we recover the SM Higgs sector.
The choice of $\Lambda_H$ is subject to the condition $\Lambda_H^2 \gg v^2$, so that we are safely in the decoupling regime.
Moreover, in order for the 2HDM to be distinguishable from the SM Higgs sector, $\Lambda_H$ should not be significantly larger than $\mathcal{O}(1~{\rm TeV})$.
We considered two different values, $\Lambda_H =$ 500 GeV and 1 TeV, although the allowed parameter regime in which the 2HDM remains consistent up to the Planck scale is not especially sensitive to the precise value of $\Lambda_H$ in the desired mass range.
In the case of $\Lambda_H=500$~GeV, it is plausible that the heavy Higgs boson states could be detected in high luminosity LHC running.  Indeed, as we shall demonstrate later in this section, differences in the squared-masses of the heavy Higgs states can provide an important consistency check of this framework.

\begin{figure}[t!]
\includegraphics[width=0.5\textwidth]{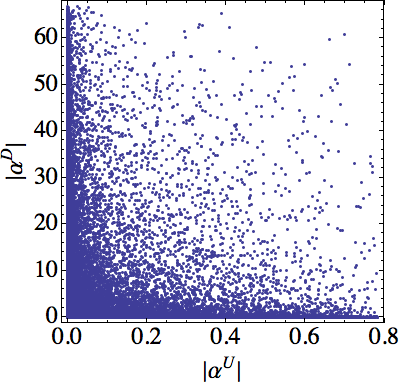}
\caption{\small Distribution of absolute values of the flavor alignment parameters,
for regions of 2HDM parameter
space which remain valid up to the Planck scale, assuming $\Lambda_H=500$~GeV.}
\label{fig:aligns}
\end{figure}

The Yukawa couplings play a fundamental role in this analysis.  As discussed in
Section~\ref{sec:Yukawa}, we have employed the flavor aligned 2HDM to describe the Yukawa sector,
with random complex alignment parameters whose moduli were varied by several orders of magnitude.
The evolution of the Yukawa couplings in the flavor-aligned 2HDM was first performed in Ref.~\cite{Bijnens:2011gd}.
Notice that the running of the Yukawa couplings can also generate Landau poles.  Due to the large size of
the top quark mass, at least one of the Yukawa couplings will be of order one at the electroweak scale,
so that a Landau pole in the top-quark Yukawa coupling below the Planck scale can be generated by the RG running.
The alignment parameters, unique for both the up and down quark sectors, were log random generated in such a 
way as to prevent such Landau poles in the running of the Yukawa couplings up to Planck scale.  In the RG running,
the initial
value of the top Yukawa coupling was taken to be $y_t (m_t)= 0.94$, corresponding to an
$\overline{\rm MS}$ top
quark mass of $m_t(m_t)=163.71 \pm 0.9$ GeV~\cite{pdg}. The non-occurrence of Landau poles then leads 
to the constraints\footnote{For $\Lambda_H=1$~TeV, we find $|\alpha^U| \lesssim 0.97$ and $|\alpha^D| \lesssim 84$.  The figure corresponding
to Fig.~\ref{fig:aligns} looks nearly identical, so we do not display it here.}
\beq
|\alpha^U| \lesssim 0.95\quad \text{and}\quad |\alpha^D| \lesssim 81.5,
\eeq
as seen in Fig. \ref{fig:aligns}.  These results are quite consistent with those obtained in Ref.~\cite{Bijnens:2011gd}.

The effect of the alignment parameters in the one-loop quartic scalar coupling RGEs is to bolster the negative
Yukawa terms, thereby further driving the quartic scalar couplings to be negative during RGE evolution.
The influence of the Yukawa couplings in the scalar couplings RG evolution is
dominated by $y_t^4$ terms (where $y_t$ is the top quark Yukawa coupling) in the one-loop
$\beta$-functions, where they provide
a negative contribution. In this manner, the large size of the top quark Yukawa coupling tends to drive $Z_1$ negative at
large energy scales, thus provoking an instability in the potential. This will occur unless the
starting point value (at the electroweak scale) of $Z_1$
is large enough. Since $Z_1$ is directly related to the lightest
CP-even mass in the decoupling regime, requiring the stability of the scalar potential between the electroweak scale and the
Planck one therefore yields a lower bound on $m_h$. Similarly, if the initial value of $Z_1$ at the electroweak scale is too large, then a Landau pole will appear in the running of $Z_1$ below the Planck scale due to the fact that the leading $Z_i$ contributions to the $\beta$-functions of the quartic scalar couplings are positive, thereby
driving the quartic scalar couplings to larger values as the energy scale increases.
Preventing the occurrence
of Landau poles thus establishes an upper bound on $Z_1$, and thus on $m_h$.

Within the SM, these demands can only be satisfied up to the Planck scale by a rather narrow
window of Higgs boson
masses, which excludes the observed value of 125 GeV. As we shall now see, the
complexity of the 2HDM scalar potential ``opens up" that narrow window to include the
known value of the Higgs mass.

\section{\label{sec:NumericalResults}Numerical Analysis}

\subsection{Results from one-loop RG running}

\begin{figure}[t!]
\includegraphics[width=0.5\textwidth]{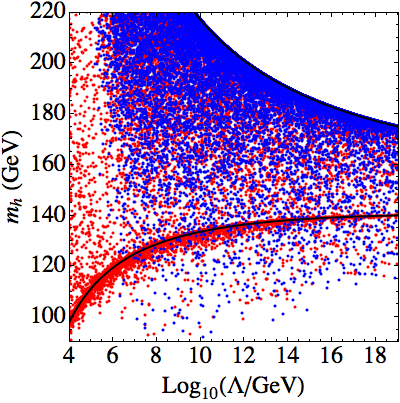}
\caption{\small RG running of 2HDM quartic scalar couplings, with $\Lambda_H=1$ TeV. Red points correspond to parameter choices for which an instability
occurs in the scalar potential; blue points indicate the presence of a Landau pole. The upper solid black line indicates the
occurrence of a Landau pole in the SM.  The lower solid black line indicates the limit for which the SM potential becomes unstable.}
\label{fig:chimneybig}
\end{figure}

Let us now compare the effect of the one-loop running of the SM scalar coupling, with its effect on the 2HDM
quartic scalar couplings. The results of our calculations are shown in Fig.~\ref{fig:chimneybig}, which we now analyze in detail.
The full 2HDM running begins at $\Lambda_H=1$~TeV, where  the $Z_i$ for $i=2,\ldots,7$ are chosen.\footnote{The corresponding plot for $\Lambda_H=500$~GeV looks nearly identical, so we do not exhibit it here.}
The red points in Fig.~\ref{fig:chimneybig} correspond to choices of parameters $Z_i$ for which an instability of the potential occurred for a given higher scale $\Lambda>\Lambda_H$.
The blue points correspond to parameter choices for which a Landau pole occurred during the RG running
at some scale $\Lambda >\Lambda_H$.  These results are to be compared with the corresponding results of the SM Higgs sector
also shown in Fig.~\ref{fig:chimneybig}: the upper solid line indicates the maximally allowed value of $m_h$ to avoid a Landau pole and the lower solid line indicates the minimal value of $m_h$ needed to avoid a negative SM quartic scalar coupling,
at all energy scales below $\Lambda$.   We recover the well-known one-loop SM
result that $140 \lesssim m_h \lesssim 175$ GeV in order to preserve vacuum stability and avoid Landau poles
in the running of the quartic scalar coupling at all energy scales up 
to $M_{\rm PL}$~\cite{Cabibbo:1979ay,Lindner:1985uk,Sher:1988mj,Lindner:1988ww,Ford:1992mv,Sher:1993mf,Altarelli:1994rb,Hambye:1996wb} .

The distribution of red and blue points in Fig.~\ref{fig:chimneybig} has some interesting features.  First, there are no blue points above the SM-Landau pole line.  In fact, although the 2HDM scalar potential
has several scalar couplings, their contributions to the 2HDM $\beta$-functions are mostly positive. As such, when one
of these couplings starts to become very large in its RG evolution, the others will not be
able to counteract that growth, and a Landau pole is reached. Consequently, the upper limit
for the quartic scalar coupling $Z_1$ that controls the value of $m_h$ hardly differs from the corresponding SM result.
Second, note the appearance of many blue points {\em below} the SM-instability line.
These correspond to Landau poles that occur for relatively low values of $m_h$, which is equivalent to low values of $Z_1$.
However, even though the initial value of $Z_1$ at $\Lambda_H$ may be small,
the values of other $Z_i$ can be large, and thus
Landau poles in these couplings can be generated, yielding those blue points below the SM instability line.

The most interesting aspect of our results concerns the distribution of the red points, which correspond to
the violation of one or more of the 2HDM stability conditions at the energy scale $\Lambda$. We see a great
``density" of points around the SM-instability line.  These points may be interpreted as regions of 2HDM
parameter space that constitute small deviations from SM behavior.
But the remarkable difference with the SM result is
the appearance of many points {\em below and to the right} of the SM-instability line.
For these points,  the instability of the scalar potential occurs at a larger value of $\Lambda$ for a given
value of $m_h$ as compared to the SM.
Indeed, the full impact of the 2HDM on the RG evolution may be best
appreciated by examining the rightmost boundary of Fig.~\ref{fig:chimneybig} corresponding to $\Lambda=M_{\rm PL}$.
On this boundary, we find both blue and red points, for a
range of Higgs masses
from about 118 GeV up to 175 GeV. Thus we see that a range of 2HDM parameters exists for which it is possible to 
have a SM-like Higgs boson with a mass of 125 GeV, without that mass value implying an instability of the 
potential (or a Landau pole) between the electroweak and Planck scales.

Let us now analyze more closely the region of parameter space for which the 2HDM is consistent up
to the Planck scale.
According to Fig.~\ref{fig:chimneybig}, only a narrow range of $m_h$  (which corresponds to
a narrow interval of values of $Z_1$) is consistent with a 2HDM with a stable vacuum and no Landau poles
from the electroweak to the Planck scale.  Since the 2HDM quartic couplings are all coupled together in their RG running,
it follows that the allowed ranges for all $Z_i$, not only $Z_1$, will likewise be quite narrow.  This has interesting
implications on the scalar mass spectrum. In fact, in light of
eq.~\ref{eqn:masssplit1}, the squared-mass splitting of the two heavy neutral Higgs states
depends primarily on $|Z_5|$.  Likewise,
eq. \ref{eqn:masssplit2} shows that the squared-mass splitting of the heavier neutral Higgs boson and
the charged Higgs boson primarily depends on $Z_4$ and $|Z_5|$. Since the possible values of $Z_4$ and $|Z_5|$ are restricted to a narrow range of values, it follows that the squared-mass splittings of the heavy Higgs states should also be strongly constrained.

\begin{figure}[t!]
\includegraphics[width=0.45\textwidth]{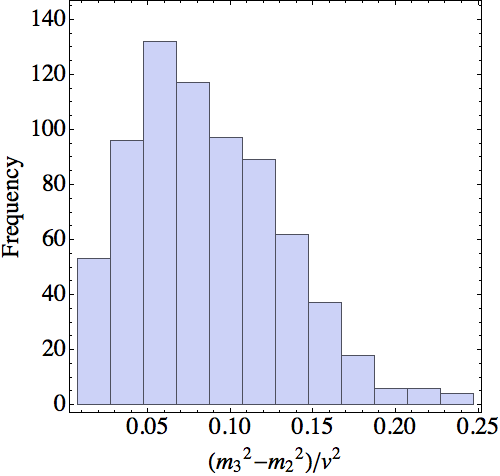}\hspace{0.3in}
\includegraphics[width=0.45\textwidth]{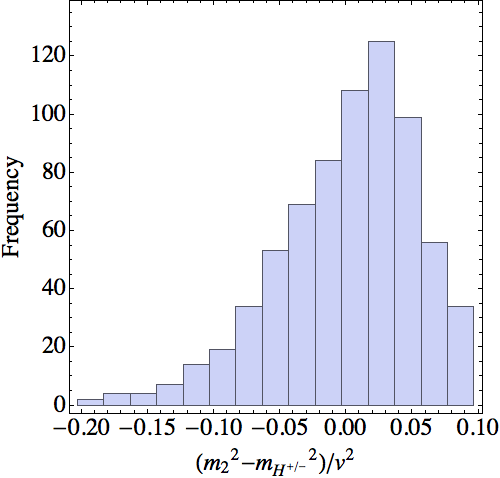}
\caption{\small Histograms of squared-mass differences of the heavy scalar states for $\Lambda_H=500$~GeV.  The left panel shows values of squared-mass difference between the two heavier neutral states.
The right panel shows the values of the squared-mass difference between the lighter of the two heavy neutral states and the charged Higgs boson.  The histograms correspond to 2HDM parameters for which there are no Landau poles and vacuum stability is satisfied at all energies below the Planck scale.}
\label{fig:histo}
\end{figure}

For a 125 GeV SM-like Higgs boson, we have evaluated the squared-mass splittings of the heavier Higgs bosons for 2HDM parameters that are consistent with a stable scalar potential and an absence of Landau poles up
to the Planck scale.  The histograms shown in Fig.~\ref{fig:histo} exhibit the distributions in arbitrary units of the squared-mass difference between the two heavy neutral states (which is positive by definition) and the difference between the lighter of the two heavy neutral states and the charged Higgs pair, for $\Lambda_H =$ 500 GeV. 
Given the formulae in section~\ref{sec:decouple}, all the heavy scalars have masses of order $\Lambda_H$ in the 
decoupling limit. 
The statistics of these histograms are summarized in Table~\ref{table:sms500}.
If the 2HDM is valid up to the Planck scale, then the mass differences among the heavy Higgs states must be quite small.   This presents a challenge for heavy Higgs searches at future colliders.   It may be that such a spectrum could only be reliably determined at a multi-TeV lepton collider.  Indeed, if
the heavy Higgs spectrum could be determined at some future collider, it would provide a
nontrivial check of the present framework in which the 2HDM is valid up to the Planck scale.

The results shown in Table~\ref{table:sms500}  are not particularly sensitive to the value of $\Lambda_H$.  For example,
if $\Lambda_H$ = 1 TeV, then the distribution of possible squared-mass differences yields the results shown in
Table~\ref{table:sms1}.   Of course, in this case the corresponding mass differences are even smaller, and the separate  discovery of each of these new scalar states at a future collider is even more challenging.
\begin{table}
\begin{tabular}{|c|| c | c | c | c |}
\hline
& min & max & mean & std. dev.  \\
\hline
$(m_3^2 - m_2^2)/v^2$ & 0.01 & 0.26 & 0.09 & 0.05 \\
$(m_2^2 - m_{H^\pm}^2)/v^2$ & $-0.20$ & 0.11 & 0 & 0.05  \\
$(m_3^2 - m_{H^\pm}^2)/v^2$ & $-0.07$ & 0.19 & 0.09 & 0.04 \\
\hline
\end{tabular}
\caption{\small Squared mass splittings of the heavier Higgs bosons of the 2HDM with $\Lambda_H$ = 500 GeV, for
$124 \lesssim m_h \lesssim 126$ GeV, for points that survive up to the Planck scale, using
 one-loop calculations.}
\label{table:sms500}
\end{table}

\begin{table}
\begin{tabular}{|c|| c | c | c | c |}
\hline
& min & max & mean & std. dev.  \\
\hline
$(m_3^2 - m_2^2)/v^2$ & 0 & 0.29 & 0.09 & 0.05 \\
$(m_2^2 - m_{H^\pm}^2)/v^2$ & $-0.23$ & 0.12 & 0 & 0.06  \\
$(m_3^2 - m_{H^\pm}^2)/v^2$ & $-0.08$ & 0.19 & 0.09 & 0.04 \\
\hline
\end{tabular}
\caption{\small Squared mass splittings of the heavier Higgs bosons of the 2HDM with $\Lambda_H$ = 1 TeV, for
$124 \lesssim m_h \lesssim 126$ GeV, for points that survive up to the Planck scale, using
 one-loop calculations.}
\label{table:sms1}
\end{table}

\subsection{The effects of two-loops RG running}

In the SM, the inclusion of the two-loop terms in the RGEs shifts the scalar potential instability boundary to a
higher energy scale, which lowers the minimum Higgs boson mass that is consistent with a stable scalar potential all the way up to the Planck scale.  In particular, the results in Ref. \cite{Degrassi:2012ry} yield a minimal value of $m_h \simeq129$ GeV for vacuum stability.  Moreover, given the currently observed value of 125 GeV for the Higgs boson, the SM vacuum is metastable under the assumption of no new physics beyond the Standard Model below about $10^{10}$~GeV.
This means that the effect of including two-loop effects in the RG running lowers by about 10~GeV the minimal  
value of the Higgs mass that is consistent with vacuum stability.

\begin{figure}[t!]
      \centering
        \begin{subfigure}[b]{0.45\textwidth}
                \centering
                \includegraphics[width=\textwidth]{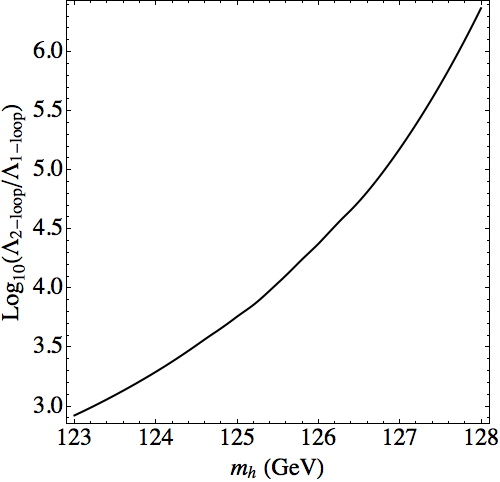}
                \label{fig:scaleshift}
        \end{subfigure}
       \centering
        \begin{subfigure}[b]{0.45\textwidth}
                \centering
                \includegraphics[width=\textwidth]{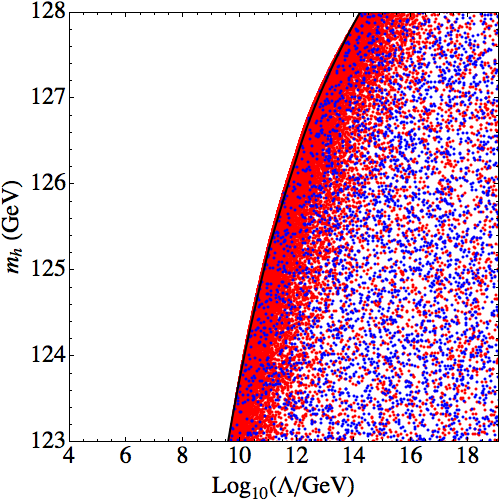}
                \label{fig:chimney2L}
        \end{subfigure}
       \caption{\small Left panel: Scale shift for converting the one-loop scalar potential instability boundary to the two-loop scalar potential instability boundary for a SM-like Higgs boson. Right panel: Higgs boson mass bounds in the flavor-aligned 2HDM, incorporating the scale shift shown in the left panel, assuming that $\Lambda_H=1$~TeV. Red points indicate an instability in the running; blue points indicate the presence of a Landau pole.}
        \label{fig:2Lshifts}
\end{figure}

\begin{table}
\begin{tabular}{|c|| c | c | c | c |}
\hline
& min & max & mean & std. dev.  \\
\hline
$(m_3^2 - m_2^2)/v^2$ & 0 & 0.31 & 0.11 & 0.05 \\
$(m_2^2 - m_{H^\pm}^2)/v^2$ & $-0.23$ & 0.12 & 0 & 0.05  \\
$(m_3^2 - m_{H^\pm}^2)/v^2$ & $-0.09$ & 0.23 & 0.11 & 0.04 \\
\hline
\end{tabular}
\caption{\small Squared-mass splittings of the heavy Higgs bosons of the 2HDM with $\Lambda_H$ = 1 TeV, for $m_h \simeq 125$ GeV, for points that survive to the Planck scale, using the two-loop extended procedure.}
\label{table:sms2}
\end{table}

We expect that employing the full two-loop RG analysis for the 2HDM would provide a similar downward shift in the
lower bound of Higgs masses that survive up to the Planck scale, as well as increase the fraction of points that
survive. In practice, implementing this full two-loop procedure is computationally
expensive. Instead, we present a procedure to estimate the two-loop RG
results. Note that the stability curve for the SM scalar potential at two-loops is both shifted to a higher
energy scale, and is less steep as a function of the Higgs mass, relative to the one-loop SM scalar potential
stability curve. In essence, going from one-loop to two-loops shifts the stability curve energy scale to a
higher scale for a particular Higgs mass. From our one-loop SM calculations and the two-loop SM calculations
of Ref. \cite{Degrassi:2012ry}, we determine the energy scale shift of the SM scalar potential stability
curves due to the inclusion of two-loop RG running.  Taking $\Lambda_H=1$~TeV,
the resulting scale shift function is shown in the left panel of
Fig. \ref{fig:2Lshifts}, which then yields our  ``two-loop" result shown in the right panel, which is obtained by
applying the scale shift to our one-loop calculation. This shift is applied only to those points in which the scalar potential became unstable, not for
points that hit a Landau pole before the Planck scale. The upper bound on the SM Higgs mass due to the absence of Landau poles does not exhibit a similar shift from one-loop to two-loop calculations.  As in the case of Fig.~\ref{fig:chimneybig}, the case of $\Lambda_H=500$~GeV yields nearly identical results.

In our one-loop calculations, only 707 (or 0.707\%) of the 100,000 points analyzed survive to the Planck scale
in the 123 GeV to 128 GeV region. With the conversion shift and a double check that they satisfy the
stability requirement for 2HDM quartic scalar coupling parameters, 1,371 more points reach the Planck scale for a total of
2,078 (or 2.078\%) at the Planck scale, an increase of 94\% relative to the one-loop results. With an
increase
in the number of points, the ``two-loop" squared mass splittings of the heavier Higgs bosons for points that
survive up to the Planck scale are given in Table \ref{table:sms2}.
Comparing Tables \ref{table:sms1} and \ref{table:sms2}, we see that there exists only slight differences in the squared-mass splittings of the heavier Higgs bosons when the approximate two-loop effects are included.
Nonetheless, the increase in the number of points for which the model remains consistent all the way up to the
Planck scale is according to what one should expect,
in light of the observation that the two-loop contributions increase the
stability of the SM potential.
Thus, this quick estimate suggests that the 2HDM parameter space corresponding to a stable scalar potential and no Landau poles in the RG running to the Planck scale is somewhat larger than the parameter regime identified in the one-loop analysis. 
In particular, given the observed Higgs mass of the 125 GeV, there exists a robust region of the parameter space for which the validity of the 2HDM and the stability of the Higgs vacuum is preserved up to the Planck scale.

%
%
%
%

\section{\label{sec:Conclusions}Conclusions}

The discovery of a SM-like Higgs boson with a mass $m_h=125$~GeV has focused attention on the validity of the Standard Model at higher energies.   Putting aside the question of the origin of the electroweak symmetry breaking (e.g., accepting the fine-tuning of parameters inherent in fixing the electroweak scale), one can ask whether the Standard Model is consistent all the way up to the Planck scale.   Refined calculations of the radiatively-corrected scalar potential suggest that the Standard Model vacuum is at best metastable (and long-lived), with a deeper vacuum located at field values near $10^{10}$~GeV, well below the Planck scale.

Adding new degrees of freedom has the potential of ameliorating the problem of an unstable vacuum.  In this paper we considered the two Higgs doublet extension of the Standard Model (2HDM) and examined the range of parameters for which the 2HDM is stable and perturbative at all energy scales below the Planck scale.  Our aim was to make the minimal number of assumptions regarding the structure of the 2HDM required by the experimental data.  Since the observed Higgs boson is SM-like (within the accuracy of the limited Higgs data set), we considered the 2HDM with the most general scalar potential
in the decoupling regime.
The Yukawa sector was treated using the flavor alignment ansatz, in
which the second set of Yukawa matrices is proportional to the SM-like
set at the electroweak scale to protect against tree-level
Higgs-mediated FCNCs.   Although the flavor alignment condition is not protected by a low-energy symmetry (except in special cases, which lead to 2HDMs of Types I, II, X or Y), it provides a more general framework which at present is consistent with experimental data.

We scanned over the scalar potential parameters and the flavor alignment parameters to fix the boundary conditions at
the scale of the heavy Higgs states.  We then employed one-loop RGEs to run the 2HDM parameters up to the Planck scale, and required that no Landau poles are encountered, without generating an instability in the scalar potential.  In contrast to the Standard Model, it is possible to have a SM-like Higgs boson with a mass of 125 GeV while maintaining the validity of the 2HDM up to the Planck scale.  We also presented a scheme to estimate the effects of the RG-running at two-loops, by applying a scale shift seen in going from the one-loop SM scalar potential stability curve to the two-loop SM scalar potential stability curve.   Such effects \textit{increase} the number of points in the 2HDM parameter scan that survive Landau pole and stability requirements up to the Planck scale.

The larger range of allowed values of $m_h$ in the 2HDM (as compared with the SM) is a direct consequence of the fact that the 2HDM scalar potential contains more quartic scalar couplings than the SM, which increases the stability of the potential at all scales between the electroweak and the Planck
scale. In contrast, we observed that the theoretical upper bound on $m_h$ in the 2HDM based on
the non-existence of Landau poles up to Planck scale hardly differs from the corresponding SM
behavior.  This can be understood as follows. In the SM, the negative top Yukawa contribution in the quartic scalar coupling
$\beta$-function drives that coupling to negative values during RG running, unless its starting point is sufficiently large. In the 2HDM,  even if the initial values of {\em some} of the quartic scalar couplings are small, and even though the top quark contributions to the $\beta$-functions are still negative, other couplings are allowed to have large values, which
(in some cases) counterbalance any putative instabilities arising due to
RG running. The 2HDM scalar potential is thus comparatively more stable than that of the SM.

Finally, we have obtained bounds on the square-mass differences of the heavier Higgs bosons in the parameter regime where the 2HDM remains valid up to the Planck scale.  If the 2HDM is realized in nature, this could provide an important check of the consistency of the model.

%
%
%
%

\begin{acknowledgments}
The authors would like to acknowledge useful discussions with Patrick Draper. 
We are also grateful to Alejo Rossia for pointing out a typographical error in the 
expression for the one-loop $\beta$ function for $Z_{a\bar{b}c\bar{d}}$, which appeared in Appendix A of the published version of this paper (and was not corrected in the first erratum). H.E.H and E.S. are supported in part by U.S. Department of Energy grant number DE-FG02-04ER41286.
E.S. also acknowledges the support of a National Science Foundation Graduate Research Fellow. The work of P.M.F. is supported in part by the Portuguese
\textit{Funda\c{c}\~{a}o para a Ci\^{e}ncia e a Tecnologia} (FCT)
under contract PTDC/FIS/117951/2010, by FP7 Reintegration Grant, number PERG08-GA-2010-277025,
and by PEst-OE/FIS/UI0618/2011.
\end{acknowledgments}


%
%
%
%

\appendix
\section{One-Loop Renormalization Group Equations}
\label{sec:RGEs}

The one-loop RGEs for the SM used in this analysis are provided by
Ref. \cite{Arason:1991ic}. The 2HDM one-loop RGEs in various bases are
given in Refs. \cite{Parida:1999td, Das:2000uk, Haber:1993an,
  Ivanov:2006yq, Ferreira:2010xe}.   The
one-loop RGEs found in the literature typically assume
a 2HDM scalar potential with a $\mathbb{Z}_2$ symmetry, $\Phi_1\to \Phi_1$, $\Phi_2\to -\Phi_2$,
to avoid FCNCs and/or are explicitly
CP-conserving. Here, we present one-loop RGEs for the full 2HDM, using
a basis-independent approach and making no CP assumptions.

In a general 2HDM,
the Higgs fermion
interactions are governed by the following interaction Lagrangian:
\beq \label{ymodeliii0}
-\mathscr{L}_{\rm Y}
=\anti \qlo\, \wtil\Phi_{\abar}\eiuoa\,  \uro +\anti Q_L^0\,\Phi_a(\eidoab)^\dagger \dro
+\anti E_L^0\,\Phi_a(\eieoab)^\dagger\,\ero
+{\rm h.c.}\,,
\eeq
summed over $a,\abar=1,2$,
where $\Phi_{1,2}$ are the Higgs doublets, $\wtil\Phi_{\abar}\equiv
i\sigma_2 \Phi^*_{\abar}$,
$\qlo $ and $\elo$ are the weak isospin quark and lepton doublets,
and $\uro$, $\dro$, $\ero$ are weak isospin quark and lepton singlets.
[The right and left-handed fermion fields are defined as usual:
$\psi_{R,L}\equiv P_{R,L}\psi$, where $P_{R,L}\equiv \half(1\pm\gamma_5)$.]
Here, $\qlo $, $\elo$, $\uro $,
$\dro$, $\ero$ denote the interaction basis states, which
are vectors in the quark and lepton
flavor spaces, and $\eiuo,\eiiuo,\eido,\eiido,\eieo,\eiieo$ are $3\times 3$
matrices in quark and lepton flavor spaces.

The neutral Higgs states acquire vacuum expectation values,
\beq
\langle\Phi^0_a\rangle=\frac{v\hat{v}_a}{\sqrt{2}}\,,
\eeq
where $\hat v_a \hat v^*_{\abar}=1$ and $v=246$~GeV.  It is also convenient to define
\beq
\hat w_b\equiv \hat v^*_{\abar}\epsilon_{ab}\,,
\eeq
where $\epsilon_{12}=-\epsilon_{21}=1$ and $\epsilon_{11}=\epsilon_{22}=0$.

It is convenient to define invariant and pseudo-invariant matrix Yukawa couplings~\cite{Davidson:2005cw,Haber:2006ue},
\beq \label{kapparho}
\kappa^{F,0}\equiv \hat v^*_\abar\eta^{F,0}_a\,,\qquad\qquad
\rho^{F,0}\equiv \hat w^*_\abar\eta^{F,0}_a\,,
\eeq
where $F=U$, $D$ or $E$.
Inverting these equations yields
\beq \label{inverted}
\eta^{F,0}_a=\kappa^{F,0}\hat v_a+\rho^{F,0}\hat w_a\,.
\eeq
Note that under the U(2) transformation, $\Phi_a\to U_{a\bar{b}}\Phi_b$ [cf.~\eq{what}],
\beq \label{rhotrans}
\kappa^{F,0}~~\hbox{is invariant and}~~\rho^{F,0}\to (\det U)\rho^{F,0}\,.
\eeq

The Higgs fields in the Higgs basis are defined by~\cite{Haber:2006ue}
\beq \label{Hdef}
H_1\equiv \hat v^*_{\abar}\Phi_a\,,\qquad\quad H_2\equiv \hat w^*_{\abar}\Phi_a\,,
\eeq
which can be inverted to yield
$
\Phi_a=H_1 \hat v_a+H_2 \hat w_a\,.
$
One can rewrite \eq{ymodeliii0} in terms of the Higgs basis fields,
\beqa \label{yukhbasis}
-\mathscr{L}_{\rm Y}
&=&\anti \qlo\, (\wtil H_1\kappa^{U,0}+\wtil H_2 \rho^{U,0})\,  \uro +\anti Q_L^0\,(H_1\kappa^{D,0\,\dagger}+H_1\rho^{D,0\,\dagger})\, \dro \nonumber \\
&&\qquad\qquad +\anti E_L^0\,(H_1\kappa^{E,0\,\dagger}+H_1\rho^{E,0\,\dagger})\,\ero
+{\rm h.c.}\,,
\eeqa

The next step is to identify the quark and lepton mass-eigenstates.  This is accomplished by
replacing $H_1\to  (0\,,\,v/\sqrt{2})$ and
performing unitary transformations of the left and right-handed up and
down quark and lepton multiplets such that the resulting quark and charged lepton mass matrices are
diagonal with non-negative entries.  In more detail, we define:
\beqa \label{biunitary}
&& P_L U=V_L^U P_L U^0\,,\qquad P_R U=V_R^U P_R U^0\,,\qquad
P_L D=V_L^D P_L D^0\,,\qquad P_R D=V_R^D P_R D^0\,,\nonumber \\
&& P_L E=V_L^E P_L E^0\,,\qquad P_R E=V_R^D P_R E^0\,,\qquad
 P_L N=V_L^E P_L N^0\,,
\eeqa
and the Cabibbo-Kobayashi-Maskawa (CKM) matrix is defined as
$
K\equiv V_L^U V_L^{D\,\dagger}\,.
$
Note that for the neutrino fields, we are free to choose
$V_L^N=V_L^E$ since neutrinos are exactly massless in this analysis.  (Here we
are ignoring the right-handed neutrino sector, which gives mass to neutrinos via the seesaw mechanism).

In particular, the unitary matrices $V_L^F$ and $V_R^F$ (for $F=U$, $D$ and $E$)
are chosen such that
\beqa
M_U&=&\frac{v}{\sqrt{2}}V_L^U \kappa^{U,0} V_R^{U\,\dagger}={\rm diag}(m_u\,,\,m_c\,,\,m_t)\,,\label{MU}\\[8pt]
M_D&=&\frac{v}{\sqrt{2}}V_L^D \kappa^{D,0\,\dagger} V_R^{D\,\dagger}={\rm
diag}(m_d\,,\,m_s\,,\,m_b) \,,\label{MD}\\[8pt]
M_E&=&\frac{v}{\sqrt{2}}V_L^E \kappa^{E,0\,\dagger} V_R^{E\,\dagger}={\rm
diag}(m_e\,,\,m_\mu\,,\,m_\tau) \label{ME}\,.
\eeqa
It is convenient to define
\beqa
\kappa^{U}&=& V_L^U \kappa^{U,0} V_R^{U\,\dagger}\,,\qquad \kappa^{D}= V_R^D \kappa^{D,0} V_L^{D\,\dagger}\,,\qquad
\kappa^{E}= V_R^D \kappa^{E,0} V_L^{E\,\dagger}\,,\label{kappas} \\
\rho^{U}&=& V_L^U \rho^{U,0} V_R^{U\,\dagger}\,,\qquad \rho^{D}= V_R^D \rho^{D,0} V_L^{D\,\dagger}\,,\qquad
\rho^{E}= V_R^D \rho^{E,0} V_L^{E\,\dagger}\,.\label{rhos}
\eeqa
\Eq{rhotrans} implies that under the U(2) transformation, $\Phi_a\to U_{a\bar{b}}\Phi_b$,
$\kappa^{F}$ is invariant, whereas $\rho^{F}\to (\det U)\rho^{F}$,
for $F=U$, $D$ and $E$.   Indeed, $\kappa^F$ is invariant since \eqst{MU}{ME} imply
that
$M_F=v\kappa^F/\sqrt{2}$,
which is a physical observable.
The matrices $\rho^U$, $\rho^D$ and $\rho^E$ are independent pseudoinvariant complex $3\times 3$ matrices.
The Higgs-fermion interactions
given in \eq{yukhbasis} can be rewritten in terms of the quark and lepton mass eigenstates,
\beqa \label{yukhbasis2}
-\mathscr{L}_{\rm Y}&=&\anti U_L (\kappa^U H_1^{0\,\dagger}
+\rho^U H_2^{0\,\dagger})\ur
-\anti D_L K^\dagger(\kappa^U H_1^{-}+\rho^U H_2^{-})\ur \nonumber \\[6pt]
&& +\anti U_L K (\kappa^{D\,\dagger}H_1^++\rho^{D\,\dagger}H_2^+)\dr
+\anti D_L (\kappa^{D\,\dagger}H_1^0+\rho^{D\,\dagger}H_2^0)\dr \nonumber \\[6pt]
&& +\anti N_L  (\kappa^{E\,\dagger}H_1^++\rho^{E\,\dagger}H_2^+)\er
+\anti E_L (\kappa^{E\,\dagger}H_1^0+\rho^{E\,\dagger}H_2^0)\er
+{\rm h.c.}
\eeqa

We now write down the renormalization group equations (RGEs) for the Yukawa matrices $\eiuoa$, $\eidoa$ and $\eieoa$.
Defining $\mathcal{D} \equiv 16 \pi^2 \mu (d/d\mu)$, the RGEs are given by~\cite{Ferreira:2010xe}:
\beqa
\mathcal{D}\eiuoa&=& -\bigl(8g_s^2+\tfrac{9}{4}g^2+\tfrac{17}{12}g^{\prime\,2}\bigr) \eiuoa
+\biggl\{3{\rm Tr}\bigl[\eiuoa({\eiuobb})^\dagger+\eidoa({\eidobb})^\dagger\bigr]
+{\rm Tr}\bigl[\eieoa({\eieobb})^\dagger \bigr]\biggr\}\eiuob \nonumber \\[8pt]
&&-2(\eidobb)^\dagger\eidoa\eiuob+\eiuoa(\eiuobb)^\dagger\eiuob+\half (\eidobb)^\dagger\eidob\eiuoa+\half \eiuob(\eiuobb)^\dagger\eiuoa\,,\label{rged}\\[6pt]
\mathcal{D}\eidoa&=& -\bigl(8g_s^2+\tfrac{9}{4}g^2+\tfrac{5}{12}g^{\prime\,2}\bigr)\eidoa
+\biggl\{3{\rm Tr}\bigl[({\eidobb})^\dagger \eidoa+({\eiuobb})^\dagger \eiuoa\bigr]
+{\rm Tr}\bigl[({\eieobb})^\dagger \eieoa\bigr]\biggr\}\eidob \nonumber \\[8pt]
&&-2\eidob\eiuoa(\eiuobb)^\dagger+\eidob(\eidobb)^\dagger\eidoa+\half \eidoa\eiuob(\eiuobb)^\dagger
+\half \eidoa(\eidobb)^\dagger\eidob\,,\label{rgeu}\\[6pt]
\mathcal{D}\eieoa&=& -\bigl(\tfrac{9}{4}g^2+\tfrac{15}{4}g^{\prime\,2}\bigr)\eieoa
+\biggl\{3{\rm Tr}\bigl[({\eidobb})^\dagger \eidoa+({\eiuobb})^\dagger \eiuoa\bigr]
+{\rm Tr}\bigl[({\eieobb})^\dagger \eieoa\bigr]\biggr\}\eieob \nonumber \\[8pt]
&&+\eieob(\eieobb)^\dagger\eieoa+\half \eieoa(\eieobb)^\dagger\eieob\,.\label{rgee}
\eeqa

\begingroup
\allowdisplaybreaks
The RGEs above are true for any basis choice.  Thus, they must also be true in the Higgs basis in which $\hat v=(1,0)$ and $\hat w=(0,1)$.
In this case, we can simply choose $\eta_1^{F,0}=\kappa^{F,0}$ and $\eta_2^{F,0}=\rho^{F,0}$ to obtain the RGEs
for the $\kappa^{F,0}$ and $\rho^{F,0}$.
Alternatively, we can multiply \eqst{rged}{rgee} first by $\hat v_a^*$ and then by $\hat w_a^*$.  Expanding $\eta_{\bar a}^\dagger$,
which appears on the right-hand sides of \eqst{rged}{rgee},
in terms of $\kappa^\dagger$ and $\rho^\dagger$ using \eq{inverted}, we again obtain the RGEs
for the $\kappa^{F,0}$ and $\rho^{F,0}$.  Of course, both methods must yield the same results,
since the diagonalization matrices employed in \eqst{MU}{ME} are defined as
those that bring the mass matrices to their diagonal form at the electroweak scale.
No scale dependence is assumed in the diagonalization matrices, and as such they are
not affected by the operators~$\mathcal{D}$.
\beqa
\mathcal{D}\kappa^{U,0}&=& -\bigl(8g_s^2+\tfrac{9}{4}g^2+\tfrac{17}{12}g^{\prime\,2}\bigr) \kappa^{U,0}
+\biggl\{3{\rm Tr}\bigl[\kappa^{U,0}\kappa^{U,0\,\dagger}+\kappa^{D,0}\kappa^{D,0\,\dagger}\bigr]
+{\rm Tr}\bigl[\kappa^{E,0}\kappa^{E,0\,\dagger }\bigr]\biggr\}\kappa^{U,0} \nonumber \\[8pt]
&&+\biggl\{3{\rm Tr}\bigl[\kappa^{U,0}\rho^{U,0\,\dagger}+\kappa^{D,0}\rho^{D,0\,\dagger}\bigr]
+{\rm Tr}\bigl[\kappa^{E,0}\rho^{E,0\,\dagger }\bigr]\biggr\}\rho^{U,0}
-2\bigl(\kappa^{D,0\,\dagger}\kappa^{D,0}\kappa^{U,0}+\rho^{D,0\,\dagger}\kappa^{D,0}\rho^{U,0}\bigr)\nonumber \\[8pt]
&&+\kappa^{U,0}(\kappa^{U,0\,\dagger}\kappa^{U,0}+\rho^{U,0\,\dagger}\rho^{U,0})
+\half(\kappa^{D,0\,\dagger}\kappa^{D,0}+\rho^{D,0\,\dagger}\rho^{D,0})\kappa^{U,0}
+\half(\kappa^{U,0}\kappa^{U,0\,\dagger}+\rho^{U,0}\rho^{U,0\,\dagger})\kappa^{U,0} \,,\nonumber\\[12pt]
\mathcal{D}\rho^{U,0}&=& -\bigl(8g_s^2+\tfrac{9}{4}g^2+\tfrac{17}{12}g^{\prime\,2}\bigr) \rho^{U,0}
+\biggl\{3{\rm Tr}\bigl[\rho^{U,0}\kappa^{U,0\,\dagger}+\rho^{D,0}\kappa^{D,0\,\dagger}\bigr]
+{\rm Tr}\bigl[\rho^{E,0}\kappa^{E,0\,\dagger }\bigr]\biggr\}\kappa^{U,0} \nonumber \\[8pt]
&&+\biggl\{3{\rm Tr}\bigl[\rho^{U,0}\rho^{U,0\,\dagger}+\rho^{D,0}\rho^{D,0\,\dagger}\bigr]
+{\rm Tr}\bigl[\rho^{E,0}\rho^{E,0\,\dagger }\bigr]\biggr\}\rho^{U,0}
-2\bigl(\kappa^{D,0\,\dagger}\rho^{D,0}\kappa^{U,0}+\rho^{D,0\,\dagger}\rho^{D,0}\rho^{U,0}\bigr)\nonumber \\[8pt]
&&+\rho^{U,0}(\kappa^{U,0\,\dagger}\kappa^{U,0}+\rho^{U,0\,\dagger}\rho^{U,0})
+\half(\kappa^{D,0\,\dagger}\kappa^{D,0}+\rho^{D,0\,\dagger}\rho^{D,0})\rho^{U,0}
+\half(\kappa^{U,0}\kappa^{U,0\,\dagger}+\rho^{U,0}\rho^{U,0\,\dagger})\rho^{U,0} \,,\nonumber\\[12pt]
\mathcal{D}\kappa^{D,0}&=& -\bigl(8g_s^2+\tfrac{9}{4}g^2+\tfrac{5}{12}g^{\prime\,2}\bigr) \kappa^{D,0}
+\biggl\{3{\rm Tr}\bigl[\kappa^{D,0\,\dagger}\kappa^{D,0}+\kappa^{U,0\,\dagger}\kappa^{U,0}\bigr]
+{\rm Tr}\bigl[\kappa^{E,0\,\dagger}\kappa^{E,0}]\biggr\}\kappa^{D,0} \nonumber \\[8pt]
&& +\biggl\{3{\rm Tr}\bigl[\rho^{D,0\,\dagger}\kappa^{D,0}+\rho^{U,0\,\dagger}\kappa^{U,0}\bigr]
+{\rm Tr}\bigl[\rho^{E,0\,\dagger}\kappa^{E,0}]\biggr\}\rho^{D,0}-2(\kappa^{D,0}\kappa^{U,0}\kappa^{U,0\,\dagger}+\rho^{D,0}\kappa^{U,0}\rho^{U,0\,\dagger}) \nonumber \\[8pt]
&&+(\kappa^{D,0}\kappa^{D,0\,\dagger}+\rho^{D,0}\rho^{D,0\,\dagger})\kappa^{D,0}+\half\kappa^{D,0}(\kappa^{U,0}\kappa^{U,0\,\dagger}+\rho^{U,0}\rho^{U,0\,\dagger})
+\half\kappa^{D,0}(\kappa^{D,0\,\dagger}\kappa^{D,0}+\rho^{D,0\,\dagger}\rho^{D,0})\,,\nonumber\\[12pt]
\mathcal{D}\rho^{D,0}&=& -\bigl(8g_s^2+\tfrac{9}{4}g^2+\tfrac{5}{12}g^{\prime\,2}\bigr) \rho^{D,0}
+\biggl\{3{\rm Tr}\bigl[\kappa^{D,0\,\dagger}\rho^{D,0}+\kappa^{U,0\,\dagger}\rho^{U,0}\bigr]
+{\rm Tr}\bigl[\kappa^{E,0\,\dagger}\rho^{E,0}]\biggr\}\kappa^{D,0} \nonumber \\[8pt]
&& +\biggl\{3{\rm Tr}\bigl[\rho^{D,0\,\dagger}\rho^{D,0}+\rho^{U,0\,\dagger}\rho^{U,0}\bigr]
+{\rm Tr}\bigl[\rho^{E,0\,\dagger}\rho^{E,0}]\biggr\}\rho^{D,0}-2(\kappa^{D,0}\rho^{U,0}\kappa^{U,0\,\dagger}+\rho^{D,0}\rho^{U,0}\rho^{U,0\,\dagger}) \nonumber \\[8pt]
&&+(\kappa^{D,0}\kappa^{D,0\,\dagger}+\rho^{D,0}\rho^{D,0\,\dagger})\rho^{D,0}+\half\rho^{D,0}(\kappa^{U,0}\kappa^{U,0\,\dagger}+\rho^{U,0}\rho^{U,0\,\dagger})
+\half\rho^{D,0}(\kappa^{D,0\,\dagger}\kappa^{D,0}+\rho^{D,0\,\dagger}\rho^{D,0})\,,\nonumber\\[12pt]
\mathcal{D}\kappa^{E,0}&=& -\bigl(\tfrac{9}{4}g^2+\tfrac{15}{4}g^{\prime\,2}\bigr) \kappa^{E,0}
+\biggl\{3{\rm Tr}\bigl[\kappa^{D,0\,\dagger}\kappa^{D,0}+\kappa^{U,0\,\dagger}\kappa^{U,0}\bigr]
+{\rm Tr}\bigl[\kappa^{E,0\,^\dagger}\kappa^{E,0}\bigr]\biggr\}\kappa^{E,0} \nonumber \\[8pt]
&&+\biggl\{3{\rm Tr}\bigl[\rho^{D,0\,\dagger}\kappa^{D,0}+\rho^{U,0\,\dagger}\kappa^{U,0}\bigr]
+{\rm Tr}\bigl[\rho^{E,0\,^\dagger}\kappa^{E,0}\bigr]\biggr\}\rho^{E,0} \nonumber \\[8pt]
&&+(\kappa^{E,0}\kappa^{E,0\,\dagger}+\rho^{E,0}\rho^{E,0\,\dagger})\kappa^{E,0}+\half\kappa^{E,0}(\kappa^{E,0\,\dagger}\kappa^{E,0}+\rho^{E,0\,\dagger}\rho^{E,0})\,,\nonumber \\[12pt]
\mathcal{D}\rho^{E,0}&=& -\bigl(\tfrac{9}{4}g^2+\tfrac{15}{4}g^{\prime\,2}\bigr) \rho^{E,0}
+\biggl\{3{\rm Tr}\bigl[\kappa^{D,0\,\dagger}\rho^{D,0}+\kappa^{U,0\,\dagger}\rho^{U,0}\bigr]
+{\rm Tr}\bigl[\kappa^{E,0\,^\dagger}\rho^{E,0}\bigr]\biggr\}\kappa^{E,0} \nonumber \\[8pt]
&&+\biggl\{3{\rm Tr}\bigl[\rho^{D,0\,\dagger}\rho^{D,0}+\rho^{U,0\,\dagger}\rho^{U,0}\bigr]
+{\rm Tr}\bigl[\rho^{E,0\,^\dagger}\rho^{E,0}\bigr]\biggr\}\rho^{E,0} \nonumber \\[8pt]
&&+(\kappa^E\kappa^{E,0\,\dagger}+\rho^{E,0}\rho^{E,0\,\dagger})\rho^{E,0}+\half\rho^{E,0}(\kappa^{E,0\,\dagger}\kappa^{E,0}+\rho^{E,0\,\dagger}\rho^{E,0})\,.\nonumber
\eeqa
\endgroup

\noindent
\begingroup
\allowdisplaybreaks
Using \eqs{kappas}{rhos}, we immediately obtain the RGEs for the $\kappa^F$ and $\rho^F$,
\beqa
\mathcal{D}\kappa^U&=& -\bigl(8g_s^2+\tfrac{9}{4}g^2+\tfrac{17}{12}g^{\prime\,2}\bigr) \kappa^U
+\biggl\{3{\rm Tr}\bigl[\kappa^U\kappa^{U\,\dagger}+\kappa^D\kappa^{D\,\dagger}\bigr]
+{\rm Tr}\bigl[\kappa^E\kappa^{E\,\dagger }\bigr]\biggr\}\kappa^U \nonumber \\[8pt]
&+&\biggl\{3{\rm Tr}\bigl[\kappa^U\rho^{U\,\dagger}+\kappa^D\rho^{D\,\dagger}\bigr]
+{\rm Tr}\bigl[\kappa^E\rho^{E\,\dagger }\bigr]\biggr\}\rho^U
-2K\bigl(\kappa^{D\,\dagger}\kappa^D K^\dagger\kappa^U+\rho^{D\,\dagger}\kappa^D K^\dagger\rho^U\bigr)\nonumber \\[8pt]
&+&\kappa^U(\kappa^{U\,\dagger}\kappa^U+\rho^{U\,\dagger}\rho^U)
+\half K (\kappa^{D\,\dagger}\kappa^D+\rho^{D\,\dagger}\rho^D)K^\dagger\kappa^U
+\half(\kappa^U\kappa^{U\,\dagger}+\rho^U\rho^{U\,\dagger})\kappa^U \,,\label{rge1}\\[12pt]
\mathcal{D}\rho^U&=& -\bigl(8g_s^2+\tfrac{9}{4}g^2+\tfrac{17}{12}g^{\prime\,2}\bigr) \rho^U
+\biggl\{3{\rm Tr}\bigl[\rho^U\kappa^{U\,\dagger}+\rho^D\kappa^{D\,\dagger}\bigr]
+{\rm Tr}\bigl[\rho^E\kappa^{E\,\dagger }\bigr]\biggr\}\kappa^U \nonumber \\[8pt]
&+&\biggl\{3{\rm Tr}\bigl[\rho^U\rho^{U\,\dagger}+\rho^D\rho^{D\,\dagger}\bigr]
+{\rm Tr}\bigl[\rho^E\rho^{E\,\dagger }\bigr]\biggr\}\rho^U
-2K\bigl(\kappa^{D\,\dagger}\rho^D K^\dagger\kappa^U+\rho^{D\,\dagger}\rho^D K^\dagger\rho^U\bigr)\nonumber \\[8pt]
&+&\rho^U(\kappa^{U\,\dagger}\kappa^U+\rho^{U\,\dagger}\rho^U)
+\half K (\kappa^{D\,\dagger}\kappa^D+\rho^{D\,\dagger}\rho^D)K^\dagger\rho^U
+\half(\kappa^U\kappa^{U\,\dagger}+\rho^U\rho^{U\,\dagger})\rho^U \,,\label{rge2}\\[12pt]
\mathcal{D}\kappa^D&=& -\bigl(8g_s^2+\tfrac{9}{4}g^2+\tfrac{5}{12}g^{\prime\,2}\bigr) \kappa^D
+\biggl\{3{\rm Tr}\bigl[\kappa^{D\,\dagger}\kappa^D+\kappa^{U\,\dagger}\kappa^U\bigr]
+{\rm Tr}\bigl[\kappa^{E\,\dagger}\kappa^E]\biggr\}\kappa^D \nonumber \\[8pt]
&+&\biggl\{3{\rm Tr}\bigl[\rho^{D\,\dagger}\kappa^D+\rho^{U\,\dagger}\kappa^U\bigr]
+{\rm Tr}\bigl[\rho^{E\,\dagger}\kappa^E]\biggr\}\rho^D-2(\kappa^D K^\dagger \kappa^U\kappa^{U\,\dagger}+\rho^D K^\dagger \kappa^U\rho^{U\,\dagger})K \nonumber \\[8pt]
&+&(\kappa^D\kappa^{D\,\dagger}+\rho^D\rho^{D\,\dagger})\kappa^D+\half\kappa^D K^\dagger(\kappa^U\kappa^{U\,\dagger}+\rho^U\rho^{U\,\dagger})K
+\half\kappa^D(\kappa^{D\,\dagger}\kappa^D+\rho^{D\,\dagger}\rho^D)\,,\label{rge3}\\[12pt]
\mathcal{D}\rho^D&=& -\bigl(8g_s^2+\tfrac{9}{4}g^2+\tfrac{5}{12}g^{\prime\,2}\bigr) \rho^D
+\biggl\{3{\rm Tr}\bigl[\kappa^{D\,\dagger}\rho^D+\kappa^{U\,\dagger}\rho^U\bigr]
+{\rm Tr}\bigl[\kappa^{E\,\dagger}\rho^E]\biggr\}\kappa^D \nonumber \\[8pt]
&+&\biggl\{3{\rm Tr}\bigl[\rho^{D\,\dagger}\rho^D+\rho^{U\,\dagger}\rho^U\bigr]
+{\rm Tr}\bigl[\rho^{E\,\dagger}\rho^E]\biggr\}\rho^D-2(\kappa^D K^\dagger \rho^U\kappa^{U\,\dagger}+\rho^D K^\dagger \rho^U\rho^{U\,\dagger})K \nonumber \\[8pt]
&+&(\kappa^D\kappa^{D\,\dagger}+\rho^D\rho^{D\,\dagger})\rho^D+\half\rho^D K^\dagger(\kappa^U\kappa^{U\,\dagger}+\rho^U\rho^{U\,\dagger})K
+\half\rho^D(\kappa^{D\,\dagger}\kappa^D+\rho^{D\,\dagger}\rho^D)\,,\label{rge4}\\[12pt]
\mathcal{D}\kappa^E&=& -\bigl(\tfrac{9}{4}g^2+\tfrac{15}{4}g^{\prime\,2}\bigr) \kappa^E
+\biggl\{3{\rm Tr}\bigl[\kappa^{D\,\dagger}\kappa^D+\kappa^{U\,\dagger}\kappa^U\bigr]
+{\rm Tr}\bigl[\kappa^{E\,^\dagger}\kappa^E\bigr]\biggr\}\kappa^E+\biggl\{3{\rm Tr}\bigl[\rho^{D\,\dagger}\kappa^D \nonumber \\[8pt]
&+&\rho^{U\,\dagger}\kappa^U\bigr]
+{\rm Tr}\bigl[\rho^{E\,^\dagger}\kappa^E\bigr]\biggr\}\rho^E+(\kappa^E\kappa^{E\,\dagger}+\rho^E\rho^{E\,\dagger})\kappa^E+\half\kappa^E(\kappa^{E\,\dagger}\kappa^E+\rho^{E\,\dagger}\rho^E)\,,\label{rge5} \\[12pt]
\mathcal{D}\rho^E&=& -\bigl(\tfrac{9}{4}g^2+\tfrac{15}{4}g^{\prime\,2}\bigr) \rho^E
+\biggl\{3{\rm Tr}\bigl[\kappa^{D\,\dagger}\rho^D+\kappa^{U\,\dagger}\rho^U\bigr]
+{\rm Tr}\bigl[\kappa^{E\,^\dagger}\rho^E\bigr]\biggr\}\kappa^E +\biggl\{3{\rm Tr}\bigl[\rho^{D\,\dagger}\rho^D\nonumber \\[8pt]
&+&\rho^{U\,\dagger}\rho^U\bigr]
+{\rm Tr}\bigl[\rho^{E\,^\dagger}\rho^E\bigr]\biggr\}\rho^E+(\kappa^E\kappa^{E\,\dagger}+\rho^E\rho^{E\,\dagger})\rho^E+\half\rho^E(\kappa^{E\,\dagger}\kappa^E+\rho^{E\,\dagger}\rho^E)\,.\label{rge6}
\eeqa
\endgroup

The 2HDM scalar potential in a generic basis shown in \eq{eq:generic} can be written in a more compact form following the notation of Ref.~\cite{Davidson:2005cw},
\begin{equation} \label{calv}
{\cal{V}} = Y_{a\bar{b}} (\Phi^{\dagger}_{\bar{a}} \Phi_b) +\tfrac{1}{2} Z_{a\bar{b}c\bar{d}}(\Phi^{\dagger}_{\bar{a}} \Phi_b) (\Phi^{\dagger}_{\bar{c}} \Phi_d)\,.
\end{equation}
Hermiticity requires that $Y_{a\bar{b}} = Y_{b\bar{a}}^*$ and  $Z_{a\bar{b}c\bar{d}} = Z_{d\bar{c}b\bar{a}}^*$.  In addition, the form of the scalar potential given in \eq{calv} implies that $Z_{a\bar{b}c\bar{d}} = Z_{c\bar{d}a\bar{b}}$.  The full one-loop $\beta$-function for $Z_{a\bar{b}c\bar{d}}$ is given by,
\beqa
&&
\mathcal{D}Z_{a\bar{b}c\bar{d}} =
4Z_{a\bar{b}e\bar{f}}Z_{c\bar{d}f\bar{e}}+
2Z_{a\bar{f}c\bar{e}}Z_{f\bar{b}e\bar{d}}+
2Z_{a\bar{f}e\bar{d}}Z_{f\bar{b}c\bar{e}}+
2Z_{a\bar{b}e\bar{f}}Z_{c\bar{e}f\bar{d}}+
2Z_{a\bar{e}f\bar{b}}Z_{c\bar{d}e\bar{f}}\nonumber \\[8pt]
&&\qquad  + \tfrac{3}{4}\bigl(3g^4-2g^{\prime 2}g^2 +g^{\prime 4}\bigr)\delta_{a\bar{b}}\delta_{c\bar{d}}
+ 3g^{\prime 2}g^2\delta_{a\bar{d}}\delta_{c\bar{b}}-4N_c\textrm{Tr}\bigl[\eta_a^F\eta^{F\dagger}_{\bar{b}}\eta_c^F\eta^{F\dagger}_{\bar{d}}\bigr] \nonumber \\[8pt]
&&\qquad -16\pi^2\bigl[\gamma_{a\bar{e}}Z_{e\bar{b}c\bar{d}}
+\gamma_{e\bar{b}}Z_{a\bar{e}c\bar{d}} + \gamma_{c\bar{e}}Z_{a\bar{b}e\bar{d}}
+ \gamma_{e\bar{d}}Z_{a\bar{b}c\bar{e}}\bigr] \,,
\label{eq:Zabcdrge}
\eeqa
where the corresponding anomalous dimension functions are given by
\beq \label{anomalousdim}
\gamma_{a\bar{b}}\equiv \half\mu\frac{d}{d\mu}\ln\bigl(\Phi_a\Phi_{\bar{b}}^\dagger\bigr)=
\frac{1}{32\pi^2}\mathcal{D}\ln\bigl(\Phi_a\Phi_{\bar{b}}^\dagger\bigr)
=\frac{1}{64\pi^2}\left\{
\bigl(3g^{\prime 2}+9g^2)\delta_{a\bar{b}}
-4N_c \text{Tr}\bigl[\eta^{F}_a\eta^{F\dagger}_{\bar{b}}\bigr]\right\}\,.
\eeq
In the notation employed in \eqs{eq:Zabcdrge}{anomalousdim}, $N_c=3$ for $F=U,D$ and $N_c=1$ for $F=E$, with an implicit sum over the repeated index $F$.  For example,
\beq
N_c \text{Tr}\bigl[\eta^{F}_a\eta^{F\dagger}_{\bar{b}}\bigr]\equiv
3 \text{Tr}\bigl[\eta^{U}_a\eta^{U\dagger}_{\bar{b}}\bigr]
+3 \text{Tr}\bigl[\eta^{D}_a\eta^{D\dagger}_{\bar{b}}\bigr]
+\text{Tr}\bigl[\eta^{E}_a\eta^{E\dagger}_{\bar{b}}\bigr]\,.
\eeq


The squared-mass and coupling coefficients of the 2HDM scalar potential in the Higgs basis [cf.~\eq{eq:HiggsBasis}] can be written in the form of invariants or pseudoinvariants with respect to the U(2) transformations, $\Phi_a\to U_{a\bar{b}}\Phi_b$, as shown in Ref.~\cite{Davidson:2005cw}.  The three squared-mass parameters are given by
\beq
Y_1 \equiv  Y_{a\bbar}\,\widehat v_\abar^\ast\, \widehat v_b\,,
\qquad\qquad
Y_2 \equiv Y_{a\bbar}\,\widehat w_\abar^\ast\, \widehat w_b\,,
\qquad\qquad
Y_3 \equiv  Y_{a\bbar}\,\widehat v_\abar^\ast\, \widehat w_b
\,,\label{yvv}
\eeq
and seven coupling parameters are given by
\clearpage

\beqa
Z_1\equiv Z_{a\bbar c\dbar}\,\widehat v_\abar^\ast\, \widehat v_b\,
\widehat v_\cbar^\ast\,\widehat v_d\,,\phantom{i}\label{zvv1}& ~~~~~~~&
Z_2 \equiv Z_{a\bbar c\dbar}\,\widehat w_\abar^\ast\, \widehat w_b\,
\widehat w_\cbar^\ast\,\widehat w_d\,,\\
Z_3 \equiv Z_{a\bbar c\dbar}\,\widehat v_\abar^\ast\, \widehat v_b\,
\widehat w_\cbar^\ast\,\widehat w_d\,,\label{zvv3}&~~~~~~~&
Z_4 \equiv
Z_{a\bbar c\dbar}\, \widehat w_\abar^\ast\, \widehat v_b\,
\widehat v_\cbar^\ast\,\widehat w_d\,,\label{zvv4} \\
Z_5 \equiv
Z_{a\bbar c\dbar}\,\widehat v_\abar^\ast\, \widehat w_b\,
\widehat v_\cbar^\ast\, \widehat w_d
\,,\label{zvv5} &~~~~~~~&
Z_6 \equiv  Z_{a\bbar c\dbar}\,\widehat v_\abar^\ast\,\widehat v_b\,
\widehat v_\cbar^\ast\, \widehat w_d\,,\label{zvv6}\\
&& \hspace{-1in} Z_7 \equiv
     Z_{a\bbar c\dbar}\,\widehat v_\abar^\ast\, \widehat w_b\,
\widehat w_\cbar^\ast\,\widehat w_d\,.\label{zvv7}
\eeqa
Note that under a U(2) transformation, $\widehat v_a\to
U_{a\bbar}\widehat v_b$, whereas
\beq \label{what}
\widehat w_a\to (\det U)^{-1} U_{a\bbar} \widehat w_b\,.
\eeq
Consequently, $Y_1$, $Y_2$, $Z_{1,2,3,4}$ are real U(2)-invariants, whereas $Y_3$, $Z_{5,6,7}$ are potentially complex
U(2)-pseudoinvariants, which are rephased under a U(2) transformation,
\beq \label{tpseudo}
[Y_3, Z_6, Z_7]\to (\det U)^{-1}[Y_3, Z_6, Z_7]\qquad {\rm and} \qquad
Z_5\to  (\det U)^{-2} Z_5 \,.
\eeq

\begingroup
\allowdisplaybreaks
As previously noted, one can evaluate \eqs{eq:Zabcdrge}{anomalousdim} in the Higgs basis by setting $\hat v=(1,0)$ and $\hat w=(0,1)$ to obtain the RGEs for the quartic scalar couplings $Z_i$,
\beqa
\mathcal{D}Z_1 &=& 12Z_1^2+4Z_3^2+4Z_3Z_4+2Z_4^2+2|Z_5|^2+24|Z_6|^2-\bigl(3g^{\prime 2}+9g^2\bigr)Z_1+\tfrac{3}{4}\bigl(g^{\prime 4}+2g^{\prime 2}g^2+3g^4\bigr) \nonumber \\[8pt]
&&-4N_c\textrm{Tr}\bigl[\kappa^{F}\kappa^{F\dagger}\kappa^{F}\kappa^{F\dagger}\bigr]+2N_c\bigl(2\textrm{Tr}\bigl[\kappa^{F}\kappa^{F\dagger}\bigr]Z_1+\textrm{Tr}\bigl[\rho^{F}\kappa^{F\dagger}\bigr]Z_6+\textrm{Tr}\bigl[\kappa^{F}\rho^{F\dagger}\bigr]Z_6^*\bigr)  \,, \label{Z1rge}\\[12pt]  
\mathcal{D}Z_2 &=& 12Z_2^2 + 4Z_3^2 + 4Z_3Z_4+2Z_4^2+ 2|Z_5|^2+24|Z_7|^2 -\bigl(3g^{\prime 2}+9g^2\bigr)Z_2
+\tfrac{3}{4}\bigl(g^{\prime 4}+2g^{\prime 2}g^2+3g^4\bigr)\nonumber \\[8pt]
&& -4N_c\textrm{Tr}\bigl[\rho^{F}\rho^{F\dagger}\rho^{F}\rho^{F\dagger}\bigr] +2N_c\bigl(2\textrm{Tr}\bigl[\rho^{F}\rho^{F\dagger}\bigr]Z_2
+\textrm{Tr}\bigl[\rho^{F}\kappa^{F\dagger}\bigr]Z_7+\textrm{Tr}\bigl[\kappa^{F}\rho^{F\dagger}\bigr]Z_7^*\bigr) \,, \label{Z2rge}\\[12pt]  
\mathcal{D}Z_3 &=& 2\bigl(Z_1+Z_2\bigr)\bigl(3Z_3+Z_4\bigr)+4Z_3^2+2Z_4^2+2|Z_5|^2+4|Z_6|^2+4|Z_7|^2+8Z_6Z_7^*+8Z_6^*Z_7  \nonumber \\[8pt]
&&-\bigl(3g^{\prime 2}+9g^2\bigr)Z_3+\tfrac{3}{4}\bigl(g^{\prime 4}-2g^{\prime 2}g^2+3g^4\bigr)
-4N_c\textrm{Tr}\bigl[\kappa^{F}\kappa^{F\dagger}\rho^{F}\rho^{F\dagger}\bigr]+N_c\bigl(2\textrm{Tr}\bigl[\kappa^{F}\kappa^{F\dagger}\bigr]Z_3  \nonumber \\[8pt]
&&+2\textrm{Tr}\bigl[\rho^{F}\rho^{F\dagger}\bigr]Z_3 +\textrm{Tr}\bigl[\rho^{F}\kappa^{F\dagger}\bigr]Z_6+\textrm{Tr}\bigl[\kappa^{F}\rho^{F\dagger}\bigr]Z_6^*+\textrm{Tr}\bigl[\rho^{F}\kappa^{F\dagger}\bigr]Z_7+\textrm{Tr}\bigl[\kappa^{F}\rho^{F\dagger}\bigr]Z_7^*\bigr) \,, \label{Z3rge}\\[12pt]  
\mathcal{D}Z_4 &=& 2(Z_1+Z_2)Z_4+8Z_3Z_4+4Z_4^2+8|Z_5|^2+10|Z_6|^2+10|Z_7|^2+2Z_6Z_7^*+2Z_6^*Z_7  \nonumber \\[8pt]
&&-\bigl(3g^{\prime 2}+9g^2\bigr)Z_4+
\tfrac{3}{2}g^{\prime 2}g^2-4N_c\textrm{Tr}\bigl[\kappa^{F}\rho^{F\dagger}\rho^{F}\kappa^{F\dagger}\bigr]+
N_c\bigl(2\textrm{Tr}\bigl[\kappa^{F}\kappa^{F\dagger}\bigr]Z_4+2\textrm{Tr}\bigl[\rho^{F}\rho^{F\dagger}\bigr]Z_4   \nonumber \\[8pt]
&&+\textrm{Tr}\bigl[\rho^{F}\kappa^{F\dagger}\bigr]Z_6+\textrm{Tr}\bigl[\kappa^{F}\rho^{F\dagger}\bigr]Z_6^*+\textrm{Tr}\bigl[\rho^{F}\kappa^{F\dagger}\bigr]Z_7+\textrm{Tr}\bigl[\kappa^{F}\rho^{F\dagger}\bigr]Z_7^*\big) \,, \label{Z4rge}\\[12pt]  
\mathcal{D}Z_5 &=& 2Z_5\bigl(Z_1+Z_2+4Z_3+6Z_4\bigr)+10Z_6^2+10Z_7^2+4Z_6Z_7-\bigl(3g^{\prime 2}+9g^2\bigr)Z_5 \nonumber \\[8pt]
&& -4N_c\textrm{Tr}\bigl[\kappa^{F}\rho^{F\dagger}\kappa^{F}\rho^{F\dagger}\bigr]  
+2N_c\bigl(\textrm{Tr}\bigl[\kappa^{F}\kappa^{F\dagger}\bigr]Z_5+\textrm{Tr}\bigl[\rho^{F}\rho^{F\dagger}\bigr]Z_5  \nonumber \\[8pt]
&&
+\textrm{Tr}\bigl[\kappa^{F}\rho^{F\dagger}\bigr]Z_6+\textrm{Tr}\bigl[\kappa^{F}\rho^{F\dagger}\bigr]Z_7\bigr) \,, \label{Z5rge}\\[12pt]  
\mathcal{D}Z_6 &=& 12Z_1Z_6+6Z_3\bigl(Z_6+Z_7\bigr)+4Z_4\bigl(2Z_6+Z_7\bigr)+2Z_5\bigl(5Z_6^*+Z_7^*\bigr)-\bigl(3g^{\prime 2}+9g^2\bigr)Z_6   \nonumber \\[8pt]
&&-4N_c\textrm{Tr}\bigl[\kappa^{F}\kappa^{F\dagger}\kappa^{F}\rho^{F\dagger}\bigr]+N_c\bigl(3\textrm{Tr}\bigl[\kappa^{F}\kappa^{F\dagger}\bigr]Z_6+\textrm{Tr}\bigl[\rho^{F}\rho^{F\dagger}\bigr]Z_6+\textrm{Tr}\bigl[\kappa^{F}\rho^{F\dagger}\bigr]Z_1 \nonumber \\[8pt]
&&+\textrm{Tr}\bigl[\kappa^{F}\rho^{F\dagger}\bigr]Z_3 +\textrm{Tr}\bigl[\kappa^{F}\rho^{F\dagger}\bigr]Z_4+\textrm{Tr}\bigl[\rho^{F}\kappa^{F\dagger}\bigr]Z_5\bigr) \,, \label{Z6rge}\\[12pt]  
\mathcal{D}Z_7 &=& 12Z_2Z_7+ 6Z_3\bigl(Z_6+Z_7\bigr)+ 4Z_4\bigl(Z_6+2Z_7\bigr)+ 2Z_5\bigl(Z_6^*+5Z_7^*\bigr)-\bigl(3g^{\prime 2}+9g^2\bigr)Z_7  \nonumber \\[8pt]
&&-4N_c\textrm{Tr}\bigl[\rho^{F}\rho^{F\dagger}\kappa^{F}\rho^{F\dagger}\bigr]+N_c\bigl(3\textrm{Tr}\bigl[\rho^{F}\rho^{F\dagger}\bigr]Z_7+\textrm{Tr}\bigl[\kappa^{F}\kappa^{F\dagger}\bigr]Z_7+\textrm{Tr}\bigl[\kappa^{F}\rho^{F\dagger}\bigr]Z_2 \nonumber \\[8pt]
&&+\textrm{Tr}\bigl[\kappa^{F}\rho^{F\dagger}\bigr]Z_3 +\textrm{Tr}\bigl[\kappa^{F}\rho^{F\dagger}\bigr]Z_4+\textrm{Tr}\bigl[\rho^{F}\kappa^{F\dagger}\bigr]Z_5\big) \,. \label{Z7rge}
\eeqa
As previously noted, $N_c=3$ for $F=U,D$ and $N_c=1$ for $F=E$, with an implicit sum over the repeated index $F$.  
\endgroup


%
%
%
%

\section{Bounded from below conditions for a general 2HDM potential}
\label{sec:Stability}

To ensure the existence of a stable vacuum, the 2HDM scalar potential must be
bounded from below, i.e.~it must assume positive values for any
direction for which the fields are tending to infinity. This places some
restrictions on the allowed values of the quartic scalar couplings. For
the case of the scalar potential given in \eq{eq:generic} with $\lambda_6=\lambda_7=0$, those necessary and sufficient
conditions are given in \eqst{eq:cpstable1}{eq:cpstable4}.

We now review the analogous
conditions for the most general renormalizable 2HDM potential, found in Refs.~\cite{Ivanov:2006yq,Ivanov:2007de}.
It is particularly convenient to introduce a new notation for the scalar potential,
based on gauge invariant field bilinears.  Indeed, in many 2HDM studies, such as the comparison
of the value of the potential in different vacua, the classification of scalar symmetries
and stability conditions, the bilinear formalism provides a significant simplification  in the calculations.  This formalism also reveals a hidden Minkowski structure
in the potential, which was established in Refs.~\cite{Ivanov:2006yq,Ivanov:2007de}.
A similar
Minkowskian notation has been employed in Refs.~\cite{heidelberg,nishi}.

There are four independent gauge-invariant
field bilinears, which are defined by
\be
\begin{array}{rcl}
r_0 &=&
\Phi_1^\dagger \Phi_1 + \Phi_2^\dagger \Phi_2,
\\*[2mm]
r_1 &=&
-\left( \Phi_1^\dagger \Phi_2 + \Phi_2^\dagger \Phi_1 \right)
= -2\,\mbox{Re}\left( \Phi_1^\dagger \Phi_2 \right),
\\*[2mm]
r_2 &=&
 i
\left( \Phi_1^\dagger \Phi_2 - \Phi_2^\dagger \Phi_1 \right)
= -2\,\mbox{Im} \left( \Phi_1^\dagger \Phi_2 \right),
\\*[2mm]
r_3 &=&
-\left( \Phi_1^\dagger \Phi_1 - \Phi_2^\dagger \Phi_2 \right).
\end{array}
\label{eq:rs}
\ee

These four quantities form the components of a covariant four-vector,
$r_\mu=(r_0\,,\,\boldsymbol{\vec r})$  with respect to SO(3,1) transformations.  We also define
$r^\mu=g^{\mu\nu}r_\mu=(r_0\,,\,-\boldsymbol{\vec r})$ where $g^{\mu\nu}$ is the usual Minkowski metric.
It is straightforward to verify that $r_0 \ge 0$ and $r^\mu r_\mu \ge 0$, the latter being a consequence of the Schwarz inequality.  That is, the four-vector $r_\mu$ lives on or inside the
forward lightcone $LC^+$.   The vacuum that preserves SU(2)$\times$U(1) electroweak symmetry
[i.e., $\langle\Phi_1\rangle =
\langle\Phi_2\rangle = 0$] corresponds to the apex of $LC^+$; all neutral vacua correspond to the surface of $LC^+$, and any charge breaking vacua would lie on the interior of $LC^+$.
Transformations of the scalar fields that preserve the scalar field kinetic energy terms leave $r_0$ invariant and correspond to SO(3) rotations of the three-vectors, $\boldsymbol{\vec r}$.

In terms of the bilinears defined in \eq{eq:rs}, the scalar potential of eq.~\eqref{eq:HiggsBasis} can be written as

\be
{\cal V} =  -M_\mu r^\mu
+ \half r^\mu\Lambda_\mu{}^ \nu  r_\nu,
\label{eq:potr}
\ee
with the 4-vector $M_\mu$ and the mixed tensor $\Lambda_\mu{}^ \nu$ given by
\be
M_\mu = \biggl( -\half(Y_1+Y_2),\quad  \mbox{Re}~Y_3,\quad
-\mbox{Im}~Y_3 ,\quad  -\half(Y_1 - Y_2) \biggr)
\label{eq:Mmu}
\ee
and
\be
\Lambda_\mu{}^ \nu = \frac{1}{2}\left( \begin{array}{cccc}
\half(Z_1+Z_2) + Z_3 & \quad
 -\mbox{Re}\left(Z_6 + Z_7\right)  & \quad \mbox{Im}\left(Z_6 + Z_7\right) & \quad
-\half(Z_1-Z_2) \vspace{0.2cm}\\
 \mbox{Re}\left(Z_6 + Z_7\right)  & \quad
-Z_4 - \mbox{Re}~Z_5  & \quad \mbox{Im}~Z_5 & \quad
-\mbox{Re}\left(Z_6 - Z_7\right) \vspace{0.2cm}\\
-\mbox{Im}\left(Z_6 + Z_7\right) &\quad \mbox{Im}~Z_5 & \quad
-Z_4 + \mbox{Re}~Z_5 & \quad \mbox{Im}\left(Z_6 - Z_7\right) \vspace{0.2cm}\\
\half(Z_1-Z_2) & \quad
-\mbox{Re}\left(Z_6 - Z_7\right) & \quad
\mbox{Im}\left(Z_6 - Z_7\right) & \quad -\half(Z_1+Z_2) + Z_3
\end{array} \right).
\label{eq:Lambda}
\ee

To ensure that the scalar potential is bounded from below one needs to evaluate
the eigenvalues and eigenvectors of the matrix $\Lambda_\mu{}^\nu$.   Then one can determine conditions on those eigenvalues and eigenvectors such that $r^\mu\Lambda_\mu{}^ \nu  r_\nu\geq 0$.
The eigenvalues $\Lambda_a$ ($a = 0, 1, 2, 3$) of the matrix $\Lambda_\mu{}^ \nu$ will
be determined by the usual characteristic equation,
\be
\mbox{det}(\Lambda_\mu{}^\nu - \Lambda_a \,g_\mu{}^\nu) \,=\,0.
\ee
since $g_\mu{}^\nu=\delta_\mu^\nu$ is just the $4\times 4$ identity matrix. The corresponding eigenvectors corresponding to eigenvalue $\Lambda_a$ will be denoted by $V^{(a)}$.
For the most general 2HDM potential, the eigenvalues are the solutions of a quartic
equation, which can in principle be determined analytically (although the  corresponding expressions are not particularly transparent).  However, it is straightforward to numerically evaluate the eigenvalues and corresponding eigenvectors.  Note that, in general, some of the eigenvalues may be complex (since the real matrix $\Lambda_\mu{}^\nu$ is not symmetric unless $Z_6=Z_7=0$ and $Z_1=Z_2$).

Having evaluated the eigenvalues and eigenvectors of $\Lambda_\mu{}^\nu$, we make use of
{\em Proposition 10}
of Ref.~\cite{Ivanov:2006yq} to conclude that the 2HDM potential is bounded from
below {\em if and only if} the following conditions are met:
\begin{enumerate}
\item All the eigenvalues $\Lambda_a$ are real.
\item $\Lambda_0 \,> \,0$.
\item $\Lambda_0 \,> \,\{\Lambda_1\,,\,\Lambda_2\,,\,\Lambda_3\}$.  There may or may not be degeneracies among the three eigenvalues $\Lambda_i$ ($i=1,2,3$).
\item
There exist four linearly independent eigenvectors $V^{(a)}$ corresponding to the four eigenvalues
$\Lambda_a$, for $a=0,1,2,3$.
\item The eigenvector $V^{(0)} = (v_{00}, v_{10},v_{20},v_{30})$, corresponding to the
eigenvalue $\Lambda_0$, is real and time-like.  That is, it can be normalized so that
\be
|V^{(0)}|^2 \, =\, v_{00}^2 \,-\, v_{10}^2 \,-\,v_{20}^2 \,-\,v_{30}^2 \,=\,1.
\nonumber
\ee
\item The remaining three eigenvectors $V^{(i)} = (v_{0i}, v_{1i},v_{2i},v_{3i})$ are real
and space-like, i.e.~normalized so that
\be
|V^{(i)}|^2 \, =\, v_{0i}^2 \,-\, v_{1i}^2 \,-\,v_{2i}^2 \,-\,v_{3i}^2 \,=\,-1.
\nonumber
\ee
\end{enumerate}

To illustrate this technique, we shall reproduce the bounded from
below conditions for a potential with a $\mathbb{Z}_2$ symmetry in the Higgs basis so that
$Z_6=Z_7=0$.
Without loss of generality, we can choose $Z_5$ real by rephasing the Higgs basis field $H_2$.
The matrix $\Lambda = {\Lambda_\mu}^\nu$ is then given by
\be
\Lambda = \frac{1}{2}\left( \begin{array}{cccc}
\half(Z_1+Z_2) + Z_3 &
0  & 0 & -\half(Z_1-Z_2) \vspace{0.2cm}\\
0  & -Z_4 - Z_5 & 0 & 0 \vspace{0.2cm}\\
0 & 0 & -Z_4 + Z_5 & 0 \vspace{0.2cm}\\
\half(Z_1-Z_2)&
0 & 0 & -\half(Z_1+Z_2)+ Z_3
\end{array} \right),
\ee
so that two of its eigenvalues can be immediately read off as $\Lambda_1 = -Z_4 - Z_5$
and $\Lambda_2 = -Z_4 + Z_5$. The remaining two eigenvalues are
\be
\Lambda_\pm \,=\, Z_3 \pm \sqrt{Z_1 Z_2}.
\ee
Since the eigenvalues must be real, if follows that
\be
Z_1 Z_2 \,>\, 0.
\label{pZ}
\ee
$\Lambda_+$ is the largest eigenvalue and thus must corresponds to the time-like eigenvector.  Hence, we identify
$\Lambda_0 = \, Z_3 + \sqrt{Z_1 Z_2}$ and $\Lambda_3 = \, Z_3 - \sqrt{Z_1 Z_2}$.
Imposing the requirement that the scalar potential is bounded from below, it follows that the eigenvalues obtained above must all be real and obey the
following inequalities:
\begin{align}
\Lambda_0 \,> \,0  &\Rightarrow \;\;\; Z_3  >  - \sqrt{Z_1 Z_2}
\\
\Lambda_0 \,> \,\{\Lambda_1\,,\,\Lambda_2\,,\,\Lambda_3\} &\Rightarrow \;\;\;
Z_3 + Z_4 - |Z_5|   >  - \sqrt{Z_1 Z_2}\,,
\end{align}
which are the Higgs basis equivalents of  \eqs{eq:cpstable3}{eq:cpstable4}. The time-like eigenvector is $V^{(0)} = (x,0,0,y)$,
where the components $x$ and $y$ are related via the eigenvector equation by
\be
y\,=\,\frac{Z_1 + Z_2 - \sqrt{Z_1 Z_2}}{Z_1 - Z_2} \, x\,.
\ee
Since the time-like normalization condition implies that $x^2 - y^2 = 1$, we
obtain
\be
x^2\,=\,\frac{(Z_1 - Z_2)^2}{4\sqrt{Z_1 Z_2}(Z_1 + Z_2)}.
\ee
Thus we see that we must have $Z_1 + Z_2 \,>\,0$, which when combined with \eq{pZ} yields
\be
Z_1\,>\,0\;\;\; , \;\;\; Z_2 \,>\,0.
\ee
Thus we recover the Higgs basis equivalents of \eqs{eq:cpstable1}{eq:cpstable2}.

%
%
%
%


\begin{thebibliography}{99}

\bibitem{:2012gk}
  G.~Aad {\it et al.}  [ATLAS Collaboration],
  Phys.\ Lett.\ B {\bf 716}, 1 (2012)
  [arXiv:1207.7214 [hep-ex]].

\bibitem{:2012gu}
  S.~Chatrchyan {\it et al.}  [CMS Collaboration],
  Phys.\ Lett.\ B {\bf 716}, 30 (2012)
  [arXiv:1207.7235 [hep-ex]].

\bibitem{Maiani:2013nga}
  L.~Maiani, A.~D.~Polosa and V.~Riquer,
  Phys.\ Lett.\ B {\bf 724}, 274 (2013)
  [arXiv:1305.2172 [hep-ph]].

\bibitem{Arbey:2013jla}
  A.~Arbey, M.~Battaglia and F.~Mahmoudi,
  Phys.\ Rev.\ D {\bf 88}, 015007 (2013)
  [arXiv:1303.7450 [hep-ph]].

\bibitem{Cabibbo:1979ay} 
  N.~Cabibbo, L.~Maiani, G.~Parisi and R.~Petronzio,
  Nucl.\ Phys.\ B {\bf 158}, 295 (1979).

\bibitem{Lindner:1985uk} 
  M.~Lindner,
  Z.\ Phys.\ C {\bf 31}, 295 (1986).

\bibitem{Hambye:1996wb}
  T.~Hambye and K.~Riesselmann,
  Phys.\ Rev.\ D {\bf 55}, 7255 (1997)
  [hep-ph/9610272].

\bibitem{Sher:1988mj} 
  M.~Sher,
  Phys.\ Rept.\  {\bf 179}, 273 (1989).
  
\bibitem{Lindner:1988ww} 
  M.~Lindner, M.~Sher and H.~W.~Zaglauer,
  Phys.\ Lett.\ B {\bf 228}, 139 (1989).

\bibitem{Ford:1992mv}
  C.~Ford, D.~R.~T.~Jones, P.~W.~Stephenson and M.~B.~Einhorn,
  Nucl.\ Phys.\ B {\bf 395}, 17 (1993)
  [hep-lat/9210033].
  
  \bibitem{Sher:1993mf} 
  M.~Sher,
  Phys.\ Lett.\ B {\bf 317}, 159 (1993)
  [Addendum: Phys.\ Lett.\ B {\bf 331}, 448 (1994)]
  [hep-ph/9307342].
  
    \bibitem{Altarelli:1994rb}
  G.~Altarelli and G.~Isidori,
  Phys.\ Lett.\ B {\bf 337}, 141 (1994).

\bibitem{Casas:1994qy}
  J.~A.~Casas, J.~R.~Espinosa and M.~Quiros,
  Phys.\ Lett.\ B {\bf 342}, 171 (1995)
  [hep-ph/9409458];
  Phys.\ Lett.\ B {\bf 382}, 374 (1996)
  [hep-ph/9603227].

\bibitem{Degrassi:2012ry}
  G.~Degrassi, S.~Di Vita, J.~Elias-Miro, J.~R.~Espinosa, G.~F.~Giudice, G.~Isidori and A.~Strumia,
  JHEP {\bf 1208}, 098 (2012)
  [arXiv:1205.6497 [hep-ph]].

\bibitem{Bezrukov:2012sa}
  F.~Bezrukov, M.~Y.~.Kalmykov, B.~A.~Kniehl and M.~Shaposhnikov,
  JHEP {\bf 1210}, 140 (2012)
  [arXiv:1205.2893 [hep-ph]].

\bibitem{Buttazzo:2013uya}
  D.~Buttazzo, G.~Degrassi, P.~P.~Giardino, G.~F.~Giudice, F.~Sala, A.~Salvio and A.~Strumia,
  JHEP {\bf 1312}, 089 (2013)
  [arXiv:1307.3536 [hep-ph]].

  \bibitem{Espinosa:1995se}
  J.~R.~Espinosa and M.~Quiros,
  Phys.\ Lett.\ B {\bf 353}, 257 (1995)
  [hep-ph/9504241].

  \bibitem{Giudice:2008bi}
  G.~F.~Giudice,
  ``Naturally Speaking: The Naturalness Criterion and Physics at the LHC,''
  in \textit{Perspectives on LHC physics}, edited by G.~Kane and A.~Pierce (World Scientific, Singapore, 2008)
  pp.~155--178
  [arXiv:0801.2562 [hep-ph]].

  \bibitem{Wells:2013tta}
  J.~D.~Wells,
  Stud.\ Hist.\ Philos.\ Mod.\ Phys.\  {\bf 49}, 102 (2015)
  [arXiv:1305.3434 [hep-ph]].

  \bibitem{Lee:1973iz}
 T.~D.~Lee,
 Phys.\ Rev.\  D {\bf 8}, 1226 (1973).

\bibitem{Branco:2011iw}
  G.~C.~Branco, P.~M.~Ferreira, L.~Lavoura, M.~N.~Rebelo, M.~Sher and J.~P.~Silva,
  Phys.\ Rept.\  {\bf 516}, 1 (2012)
  [arXiv:1106.0034 [hep-ph]].

 \bibitem{Gonderinger:2009jp}
  M.~Gonderinger, Y.~Li, H.~Patel and M.~J.~Ramsey-Musolf,
  JHEP {\bf 1001}, 053 (2010)
  [arXiv:0910.3167 [hep-ph]].

\bibitem{Kreyerhoff:1989fa}
  G.~Kreyerhoff and R.~Rodenberg,
  Phys.\ Lett.\ B {\bf 226}, 323 (1989).

\bibitem{Freund:1992yd}
  J.~Freund, G.~Kreyerhoff and R.~Rodenberg,
  Phys.\ Lett.\ B {\bf 280}, 267 (1992).

\bibitem{Kastening:1992by}
  B.~M.~Kastening,
  hep-ph/9307224.

\bibitem{Nie:1998yn}
  S.~Nie and M.~Sher,
  Phys.\ Lett.\ B {\bf 449}, 89 (1999)
  [hep-ph/9811234].

\bibitem{Kanemura:1999xf}
  S.~Kanemura, T.~Kasai and Y.~Okada,
  Phys.\ Lett.\ B {\bf 471}, 182 (1999)
  [hep-ph/9903289].

\bibitem{Ferreira:2009jb}
  P.~M.~Ferreira and D.~R.~T.~Jones,
  JHEP {\bf 0908}, 069 (2009)
  [arXiv:0903.2856 [hep-ph]].

\bibitem{EliasMiro:2012ay}
  J.~Elias-Miro, J.~R.~Espinosa, G.~F.~Giudice, H.~M.~Lee and A.~Strumia,
  JHEP {\bf 1206}, 031 (2012)
  [arXiv:1203.0237 [hep-ph]].
  
\bibitem{Lebedev:2012zw}
  O.~Lebedev,
  Eur.\ Phys.\ J.\ C {\bf 72}, 2058 (2012)
  [arXiv:1203.0156 [hep-ph]].

  \bibitem{Pruna:2013bma}
  G.~M.~Pruna and T.~Robens,
  Phys.\ Rev.\ D {\bf 88}, 115012 (2013)
  [arXiv:1303.1150 [hep-ph]].

 \bibitem{Costa:2014qga}
  R.~Costa, A.~P.~Morais, M.~O.~P.~Sampaio and R.~Santos,
  arXiv:1411.4048 [hep-ph].

  \bibitem{Chakrabarty:2014aya} 
  N.~Chakrabarty, U.~K.~Dey and B.~Mukhopadhyaya,
  JHEP {\bf 1412}, 166 (2014)
  [arXiv:1407.2145 [hep-ph]].

\bibitem{Das:2015mwa} 
  D.~Das and I.~Saha,
  Phys.\ Rev.\ D {\bf 91}, 095024 (2015)
  [arXiv:1503.02135 [hep-ph]].

\bibitem{Chowdhury:2015yja} 
  D.~Chowdhury and O.~Eberhardt,
  arXiv:1503.08216 [hep-ph].


\bibitem{Gunion:2002zf}
  J.~F.~Gunion and H.~E.~Haber,
  Phys.\ Rev.\ D {\bf 67}, 075019 (2003)
  [hep-ph/0207010].

\bibitem{Haber:2013mia}
 H.~E.~Haber, in Proceedings of he Toyama International Workshop on Higgs as a Probe of New Physics 2013 (HPNP2013),
  Toyama, Japan, 13--16 February, 2013
  [arXiv:1401.0152 [hep-ph]].

\bibitem{Pich:2009sp}
  A.~Pich and P.~Tuzon,
  Phys.\ Rev.\ D {\bf 80}, 091702 (2009)
  [arXiv:0908.1554 [hep-ph]];
  M.~Jung, A.~Pich and P.~Tuzon,
  JHEP {\bf 1011}, 003 (2010)
  [arXiv:1006.0470 [hep-ph]].

\bibitem{Davidson:2005cw}
  S.~Davidson and H.~E.~Haber,
  Phys.\ Rev.\ D {\bf 72}, 035004 (2005)
  [Erratum-ibid.\ D {\bf 72}, 099902 (2005)]
  [hep-ph/0504050].

  \bibitem{Barger:1989fj}
  V.~D.~Barger, J.~L.~Hewett and R.~J.~N.~Phillips,
  Phys.\ Rev.\ D {\bf 41}, 3421 (1990).

  \bibitem{Aoki:2009ha}
  M.~Aoki, S.~Kanemura, K.~Tsumura and K.~Yagyu,
  Phys.\ Rev.\ D {\bf 80}, 015017 (2009)
  [arXiv:0902.4665 [hep-ph]].

\bibitem{Hall:1981bc}
  L.~J.~Hall and M.~B.~Wise,
  Nucl.\ Phys.\ B {\bf 187}, 397 (1981).

  \bibitem{Ferreira:2010xe}
  P.~M.~Ferreira, L.~Lavoura and J.~P.~Silva,
  Phys.\ Lett.\ B {\bf 688}, 341 (2010)
  [arXiv:1001.2561 [hep-ph]].

\bibitem{Branco:1999fs}
  G.~C.~Branco, L.~Lavoura and J.~P.~Silva,
  \textit{CP Violation} (Oxford University Press, Oxford, UK, 1999), Chapter 22.

\bibitem{Haber:2006ue}
  H.~E.~Haber and D.~O'Neil,
  Phys.\ Rev.\ D {\bf 74}, 015018 (2006)
  [hep-ph/0602242].

\bibitem{Haber:2010bw}
  H.~E.~Haber and D.~O'Neil,
  Phys.\ Rev.\ D {\bf 83}, 055017 (2011)
  [arXiv:1011.6188 [hep-ph]].

 \bibitem{Glashow:1976nt}
 S.~L.~Glashow and S.~Weinberg,
 Phys.\ Rev.\  D {\bf 15}, 1958 (1977).

\bibitem{Paschos:1976ay}
 E.~A.~Paschos,
 Phys.\ Rev.\  D {\bf 15}, 1966 (1977).

 \bibitem{Haber:1978jt}
  H.~E.~Haber, G.~L.~Kane and T.~Sterling,
  Nucl.\ Phys.\ B {\bf 161}, 493 (1979).

  \bibitem{Donoghue:1978cj}
  J.~F.~Donoghue and L.~F.~Li,
  Phys.\ Rev.\ D {\bf 19}, 945 (1979).
  
  \bibitem{Mahmoudi:2009zx} 
  F.~Mahmoudi and O.~Stal,
  Phys.\ Rev.\ D {\bf 81}, 035016 (2010)
  [arXiv:0907.1791 [hep-ph]].

\bibitem{seesaw}
  P.~Minkowski,
  Phys.\ Lett.\ B {\bf 67}, 421 (1977);
  M.~Gell-Mann, P.~Ramond and R.~Slansky, in \textit{Supergravity}, edited by
  D.~Freedman and P.~van Nieuwenhuizen (North Holland, Amsterdam, 1979) p.~315
  [arXiv:1306.4669 [hep-th]];
  T.~Yanagida,
  Prog.\ Theor.\ Phys.\  {\bf 64}, 1103 (1980);
  R.~N.~Mohapatra and G.~Senjanovic,
  Phys.\ Rev.\ Lett.\  {\bf 44}, 912 (1980);
  R.~N.~Mohapatra and G.~Senjanovic,
  Phys.\ Rev.\ D {\bf 23}, 165 (1981).

\bibitem{Deshpande:1977rw}
  N.~G.~Deshpande and E.~Ma,
  Phys.\ Rev.\ D {\bf 18}, 2574 (1978).


\bibitem{Ivanov:2006yq}
  I.~P.~Ivanov,
  Phys.\ Rev.\ D {\bf 75}, 035001 (2007)
  [Erratum-ibid.\ D {\bf 76}, 039902 (2007)]
  [hep-ph/0609018].


\bibitem{Chetyrkin:2000yt}
  K.~G.~Chetyrkin, J.~H.~Kuhn and M.~Steinhauser,
  Comput.\ Phys.\ Commun.\  {\bf 133}, 43 (2000)
  [hep-ph/0004189];
  B.~Schmidt and M.~Steinhauser,
  Comput.\ Phys.\ Commun.\  {\bf 183}, 1845 (2012)
  [arXiv:1201.6149 [hep-ph]].

\bibitem{pdg}
K.~A.~Olive {\it et al.}  [Particle Data Group Collaboration],
  Chin.\ Phys.\ C {\bf 38}, 090001 (2014).

\bibitem{Bijnens:2011gd}
 J.~Bijnens, J.~Lu and J.~Rathsman,
 JHEP {\bf 1205}, 118 (2012)
 [arXiv:1111.5760 [hep-ph]].

\bibitem{Arason:1991ic}
  H.~Arason, D.~J.~Castano, B.~Keszthelyi, S.~Mikaelian, E.~J.~Piard, P.~Ramond and B.~D.~Wright,
  Phys.\ Rev.\ D {\bf 46}, 3945 (1992).

\bibitem{Parida:1999td}
  M.~K.~Parida and B.~Purkayastha,
  Eur.\ Phys.\ J.\ C {\bf 14}, 159 (2000)
  [hep-ph/9902374].

\bibitem{Das:2000uk}
  C.~R.~Das and M.~K.~Parida,
  Eur.\ Phys.\ J.\ C {\bf 20}, 121 (2001)
  [hep-ph/0010004].

\bibitem{Haber:1993an}
  H.~E.~Haber and R.~Hempfling,
  Phys.\ Rev.\ D {\bf 48}, 4280 (1993)
  [hep-ph/9307201].

\bibitem{Ivanov:2007de}
 I.~P.~Ivanov,
 Phys.\ Rev.\  D {\bf 77},  015017 (2008)
 [arXiv:0710.3490 [hep-ph]].

\bibitem{heidelberg}
F. Nagel, ``New aspects of gauge-boson couplings and the Higgs sector'', Ph.D. thesis, University Heidelberg
(2004), [\texttt{http://www.ub.uni-heidelberg.de/archiv/4803]};
 M.~Maniatis, A.~von Manteuffel, O.~Nachtmann and F.~Nagel,  Eur.\ Phys.\ J.\  C {\bf 48}, 805 (2006).


\bibitem{nishi}
C.~C.~Nishi, Phys.\ Rev.\ D {\bf 74}, 036003 (2006)
[Erratum-ibid.D {\bf 76}, 119901 (2007)];
Phys.\ Rev.\ D {\bf 76}, 055013 (2007);
Phys.\ Rev.\ D {\bf 77}, 055009 (2008).

\end{thebibliography}
\end{document}